\documentclass[12pt]{article}
\usepackage[utf8]{inputenc}
\usepackage{graphicx}
\usepackage{float}
\usepackage{indentfirst}
\usepackage{cite}
\usepackage{gensymb}
\usepackage[margin=1.0in]{geometry}
\usepackage{setspace} 
\usepackage{tabularx}
\usepackage{wrapfig}
\usepackage[labelfont=bf]{caption}
\usepackage[font={footnotesize}]{caption}
\usepackage[nottoc]{tocbibind}
\usepackage{hyperref}
\usepackage{authblk}
\usepackage{amsmath}
\usepackage[table, dvipsnames]{xcolor}
\hypersetup{
    colorlinks,
    citecolor=black,
    filecolor=black,
    linkcolor=black,
    urlcolor=black
}
\usepackage{subfig}
\newcolumntype{Y}{>{\centering\arraybackslash}X}

\title{Photo-Switchable Cross-Linking in Polymer Gels: Effects on Surface Creasing and Network Relaxation during Swelling}
\author{Alyssa VanZanten, Surbhi Punhani-Schillinger, M. Reed Blocksome, Aditya Ketkar, Shih-Yuan Chen, Michelle M. Driscoll, Robert C. Ferrier, Jr., Caroline R. Szczepanski}

\begin{document}

\maketitle
\section*{Abstract}
Polymer gels with photo-responsive cross-links enable tunable mechanics and surface morphologies, making them promising for adaptive materials.
While prior work on coumarin cross-linked gels has focused on photo-mediated events in dilute solution, their network-level mechanical responses remain unclear.
Here, we design PEG hydrogels with both permanent covalent and dynamic coumarin cross-links, allowing \textit{in situ} modulation of cross-linking under wavelength specific UV light. 
Real-time FTIR and dynamic mechanical analysis show that post-cure 365 nm irradiation drives rapid dimerization, increasing storage modulus by up to 69\%, whereas cleavage of coumarin cross-links via 254 nm post-cure irradiation has a more limited effect due to attenuation in bulk samples.
Surface imaging reveals that dynamic cross-linking governs swelling-induced crease formation and evolution.
Together, these results establish design principles for hydrogels with programmable mechanics and adaptive surface topographies, advancing application in smart coatings, actuators, and responsive biomaterials.
\doublespacing
\section{Introduction}

Photo-responsive cross-linking enables \textit{in situ}, spatio-temporal control of network constraints, a valuable strategy for designing smart membranes, fouling-inhibiting coatings, targeted drug-delivery vehicles, and reprocessable plastics.
Dynamic covalent bonds that reversibly associate and dissociate under irradiation with light have been integrated into diverse polymer platforms.
Such photomediated cross-links impart self-healing and re-processing \cite{Ling2011, Lopez2014, Cazin2021, Jiang2007, Shi2024, Xiao2023, Aguirresarobe2014, Kim2015}, tunable interfacial behavior (e.g., fluorescence, wettability) \cite{Inacker2022, Anastasiadis2008, Liu2011, Liu2014, Wosnick2008}, controlled release \cite{Elchiev2023, Sinkel2010, Lin2010}, and on-demand actuation \cite{Morales2016, Zhu2021, Zhao2021, Jiang2020}.
Coumarin moieties dimerize under 365nm UV light and cleave under 254nm UV light \cite{Kabb2018, Aguirresarobe2014}, offering reversibility and efficient photoluminescence \cite{Ling2011, Ji2018}.

Gels - soft, three-dimensional cross-linked polymers - provide mobility for reactive groups while maintaining bulk dimensional stability.
Coumarin moieties have been used as gelators or primary cross-linkers in hydrogels \cite{Nagata2008, Kabb2018, Chesterman2018, Lee2012, Ji2014, Ji2018, Kim2015}, but most studies probe their photodimerization and cleavage in dilute solutions \textit{via} UV-Vis spectroscopy.
Dilution removes the topological constraints of a gel network, so such measurements cannot fully capture network-level reaction pathways and kinetics.
Additionally, attenuation of UV light in bulk samples can dramatically reduce penetration and efficiency, particularly at 254 nm. 
This means that it is especially important to consider bulk effects when assessing coumarin reaction kinetics.
Here, we study a hydrogel with \textbf{both permanent covalent and dynamic coumarin cross-links}, which ensures dimensional stability during cross-link cycling.
Bulk chemical and physical changes during coumarin dimerization and cleavage are probed by pseudo-real-time fourier transform infrared (FTIR) spectroscopy and dynamic mechanical analysis (DMA).

Cross-linking imposes network constraints essential to gel design. Cross-linking results in a three-dimensional matrix that absorbs compatible solvents, but swelling is gradual, generating stresses as solvent diffuses through the network \cite{Schott1992, Plummer2023, Daniels2017, Hong2008}. 
Property mismatches during solvent diffusion can trigger instabilities such as rupture and surface buckling \cite{Leslie2021, DeSilva2016, Wu2013, Kim2014, Xu2013, Kashihara2022, Kumar2021, Takahashi2016, Guvendiren2009, Trujillo2008, Ju2022, Tanaka, Toh2015, Hong2009}.
Most gels dissipate stresses extensively during swelling \cite{Li1990, Tanaka1979, Louf2021}, making them ideal for studying stimuli-responsive behavior in which dynamic bond exchange generates internal stresses.

Photo-sensitive dynamic bonds, among other chemistries, enable engineered responses to environmental change. For example, incorporating a light responsive protein (LOV2) and its binding partner into a polymer backbone ((GSc)3), produced a gel whose modulus and viscosity were tunable with blue light (450 to 490 nm) \cite{Duan2021}. 
This demonstrates how dynamic cross-links can be used to control stiffness and swelling kinetics \cite{Aguirresarobe2014, Lopez2014, Jiang2007, Chesterman2018, Chen1996}.
Coumarin is among the few functional groups that undergo photoreversible dimerization, with straightforward chemistry enabling functionalization and integration into diverse polymers.
Herein, we use coumarin as a secondary, dynamic cross-linker to manipulate stress dynamics during swelling.
Changes in surface topography, visualized \textit{via} optical microscopy, reveal how dimerization and cleavage affect stress dissipation.
Building on theories from self-healing polymers, we relate observed viscoelastic changes to coumarin binding efficiency and anomalous diffusion behavior.
This study is organized as follows:
Section 3.1, presents  surface crease images;
Section 3.2, compares swelling, photopolymerization, and static mechanics;
Lastly, section 3.3 analyzes chemical and mechanical changes during dimerization and cleavage using FTIR and DMA.

\section{Materials and Methods}
\subsection{Materials}
\begin{figure}[H]
    \centering
    \includegraphics[scale=0.6, angle=0, trim={0cm 7.5cm 0cm 4cm},clip]{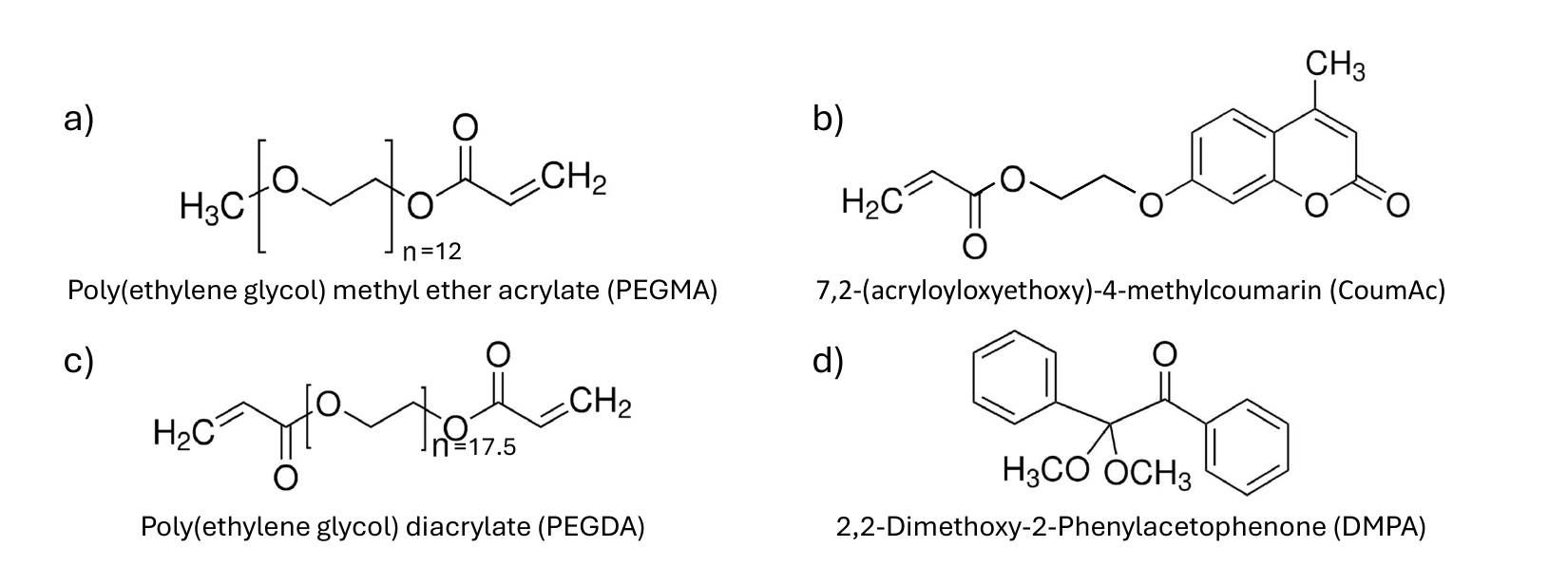}
    \caption{\textbf{Molecular structure of a) PEGMA, b) CoumAc, c) PEGDA, and d) DMPA.}}
    \label{fig:chemical structures}
\end{figure}

The gel networks investigated (shown in \textbf{Fig.\ref{fig:chemical structures}}) comprised of poly(ethylene glycol) methyl ether acrylate (PEGMA, 400 g/mol) as the monomer, poly(ethylene glycol) diacrylate (PEGDA, 700 g/mol) as the permanent cross-linker, and synthesized 7,2-(acryloyloxyethoxy)-4-methylcoumarin (CoumAc) as the photo-responsive secondary cross-linker.
2,2-dimethoxy-2-phenylacetophenone (DMPA) served as the photo-initiator in all formulations.
All chemicals were from Sigma-Aldrich and used as received, except CoumAc, which was synthesized in-house (see below).
Deionized (DI) water for swelling experiments was obtained from an in-house source.

\subsection{Methods}
    \subsubsection{Synthesis of 7,2-(acryloyloxyethoxy)-4-methylcoumarin (CoumAc)}
        CoumAc synthesis followed Kabb et al. (scheme in \textbf{Fig. S\ref{fig:rxn scheme}}) \cite{Kabb2018}. In brief, 
        4-methyl-umbelliferone (12 g) and potassium carbonate (18.82 g) were suspended in dimethylformamide under nitrogen atmosphere, then treated with 2-bromoethanol (7.2 mL). 
        The mixture was stirred at 90\degree C for 18 hours, cooled, and combined with ice cold DI water to form a pink slurry. 
        The pink/off-white precipitate was vacuum filtered, dried, and yielded 7,2-(hydroxyethoxy)-4-methylcoumarin (15 g, 100\%, 
        $^1$H-NMR in \textbf{Fig. S\ref{fig:prod1 nmr}}).
        
        The intermediate (15 g) was suspended in chloroform under nitrogen atmosphere, with triethylamine (19.2 mL) and acryloyl chloride (11 mL) added.
        After 1 hour of stirring at room temperature, additional triethylamine (9.6 mL) and acryloyl chloride (5.5 mL) were introduced, and the mixture stirred overnight. 
        The product was washed sequentially with sodium bicarbonate solution (100 mL x 2), DI water (100 mL x 2), and NaCl brine (5 M, 100 mL x 2), dried over sodium sulfate, and gravity filtered. 
        Evaporative recrystallization in ethanol (twice) afforded CoumAc (10 g, 54\%, $^1$H-NMR in \textbf{Fig. S\ref{fig:prod2 nmr}}). 
    
    \subsubsection*{Preparation and curing of PEG:CoumAc hydrogels}
    Resin formulations (\textbf{Table \ref{tab:resin formulations}}) varied the molar fractions of PEGMA, PEGDA, and, in some cases CoumAc. 
    In the cases where CoumAc was incorporated, the permanent:dynamic ratio was varied as shown in \textbf{Table \ref{tab:resin formulations}} to tune cross-link density. 
    The combined mass of these comonomers totaled 99.5 wt\% of the resin formulation.
    All formulations contained 0.5 wt\% DMPA to enable free-radical photopolymerization.

    \begin{table}[ht]
        \centering
                \caption{\textbf{Gel resin compositions and naming scheme}}

        \begin{tabular}{|c|c|c|c|}
        \hline
            Name & PEGDA mol\% & CoumAc mol\% & PEGMA mol\% \\
        \hline
            0.5:0.5:99 & 0.5 & 0.5 & 99\\
        \hline
            0.5:4.5:95 & 0.5 & 4.5 & 95\\
        \hline
            1:1:98 & 1 & 1 & 98\\
        \hline
            1:4:95 & 1 & 4 & 95\\
        \hline
            1:9:90 & 1 & 9 & 90\\
        \hline
            5:5:90 & 5 & 5 & 90\\
        \hline
            5:15:80 & 5 & 15 & 80\\
        \hline
        \end{tabular}
        \label{tab:resin formulations}
    \end{table}

    Components were combined in glass vials, stirred at 60$\celsius$ for $\sim$15 minutes to homogenize, then injected between glass slides to form 25 \textit{x} 10 \textit{x} 2 mm bars (length \textit{x} width \textit{x} thickness). 
    Photopolymerization used either: 1) irradiation with 365 nm UV LED (0.1 $\frac{W}{cm^2}$, 7 minutes, ThorLabs Solis LED 365C - this method is similar to that commonly used for photopolymerization of PEG gels, including in our prior work\cite{VanZanten2024}) or 2) 254 nm UV oven (0.0088 $\frac{W}{cm^2}$, 9-35 minutes, Stratagene UV Stratalinker 2400) with samples rotated $180\degree$ every minute to ensure a uniform cure. 
    Irradiation continued until Fourier Transform Infrared (FTIR) spectroscopy confirmed no further decreases in the vinyl peak (6165 $cm^{-1}$).
    It is important to note that the intensity values listed here are incident and do not reflect intensity values within the specimens.
    Attenuation was observed after UV light traveled through the samples, and is reported in \textbf{Table S\ref{tab:intensitydrop}}.

    \subsubsection*{Fourier Transform Infrared spectroscopy}
    
    \textbf{Tracking free-radical photopolymerization in real-time}: 
    Real-time fourier transform infrared (RT-FTIR) spectroscopy (ThermoNicolet, Nicolet iS50) monitored C=C double bond consumption during photopolymerization.
    \textit{In situ} measurements used a fiber optic UV source (250-500 nm, 0.1 $\frac{W}{(cm)^2}$ ,Omnicure series 2000m Excelitas).
    Pesudo-real-time mode paused 254 nm curing to collect spectra at discrete intervals.
    Fractional conversion was calculated from the vinyl peak at 6165 $cm^{-1}$ using Equation (1).
    \begin{equation}
        \text{Fractional Conversion}=1-\frac{A_t}{<A_0>}
    \end{equation} \label{fractional conversion}
    Here, $A_t$ is the peak area at time t and $<A_0>$ is the average peak area before irradiation.
    The rate of photopolymerization ($R_p$) was calculated as the first derivative of the fractional conversion versus time data, and maximum rate ($R_p^{max}$) and the time at $R_p^{max}$ were extracted as characteristic descriptions of the photopolymerization reaction.
    
    \textbf{Monitoring CoumAc dimerization and cleavage}: 
    Pseudo-real-time FTIR in the mid-IR range (400-4000 $cm^{-1}$, resolution of 4 $cm^{-1}$) was used to monitor the chemical changes associated with CoumAc dimerization and cleavage.
    0.01 mm thick, free-standing films were formed to mitigate saturation in the mid-IR range.
    The aromatic C=C double bond at the 3,4-position of the CoumAc moiety has a vibration at approximately 1615 $cm^{-1}$. 
    This signal is reported to appear as a sharp peak in the cleaved state \cite{Inacker2022}. 
    Dimerization is reported to reduce the intensity of the C=C double bond peak and form a shoulder at approximately 1575 $cm^{-1}$ \cite{Inacker2022}.
    The C=O stretch signal around 1730 $cm^{-1}$, which has a small shoulder at approximately 1775 $cm^{-1}$, is reportedly influenced by the C=C double bond that is present only in the cleaved state \cite{Inacker2022}. 
    Therefore, dimerization is expected to shift the C=O peak to higher wavenumbers. 
    Our observation of these chemical changes during the dimerization and cleavage reactions are included in Section 3.3.

    \subsubsection*{Dynamic mechanical analysis}
    Dynamic mechanical analysis (DMA 850, TA instruments) measured the storage modulus (\(E'\)), loss modulus (\(E"\)), and loss factor (tan \(\delta\)) during temperature sweeps (-60 or -40\degree C to 100\degree C, 1Hz, 0.01\% strain,3\degree C/min) in tensile mode.
    These measurements represent the elastic behavior of the material, the viscous behavior, and the ratio of \(\frac{E"}{E'}\), respectively.
    Strain ramp tests measured stress-strain behavior and Young's modulus (room temperature, 0.2 $\frac{mm}{min}$) in tensile mode.
    
    \subsubsection*{Hydrogel swelling and morphology measurements}
    To quantify the uptake of solvent during gel swelling, mass measurements were collected at various time points. 
    Dry gel mass ($m_0$) was recorded before immersion in 75 mL of DI water. 
    At set times, gels were removed, blotted, weighed ($m(t)$), and returned to the water. 
    Equilibrium swelling mass ($m_{eq}$) was reached when weight stabilized (typically after $\sim$ 24 hours).
    Mass swelling ratio ($Q(t)$) and normalized swelling ratio ($\overline{Q(t)}$) were calculated \textit{via} Equations \ref{swelling ratio}-\ref{norm q}.
\begin{equation} \label{swelling ratio}
    Q(t)=\frac{m(t)}{m_0}, Q_0=\frac{m_0}{m_0}=1, Q_{eq}=\frac{m_{eq}}{m_0}
\end{equation}
\begin{equation} \label{norm q}
    \overline{Q(t)}=\frac{Q(t)-Q_0}{Q_{eq}-Q_0}=\frac{m(t)-m_0}{m_{eq}-m_0}
\end{equation}
    $\overline{Q(t)}=0$ at time $t=0$ because $Q(t)=Q_0$. Additionally, $\overline{Q(t)}=1$ at equilibrium because $Q(t)=Q_{eq}$. 
    Therefore, $\overline{Q(t)}$, which increases as swelling progresses, represents the fraction of the total swelling capacity reached after time ($t$) of swelling.

    Morphology was imaged \textit{in situ} using an Olympus Microscope ix83 at 2x (6656 $\mu$m field).
    Samples were suspended and gently weighted down to prevent movement (\textbf{Fig. S\ref{fig:microscope setup}}).
    During simultaneous UV irradiation and crease imaging, a 254 nm lamp (Transilluminator Handheld UV Lamp BioGlow® 254nm) was placed $\sim5$ cm from the swelling gel, with an approximate intensity of 0.008 $W/cm^2$.

\section{Results and Discussion}
\begin{figure}[H]
    \centering
    \includegraphics[scale=0.55, angle=0, trim={0cm 4cm 0cm 3cm},clip]{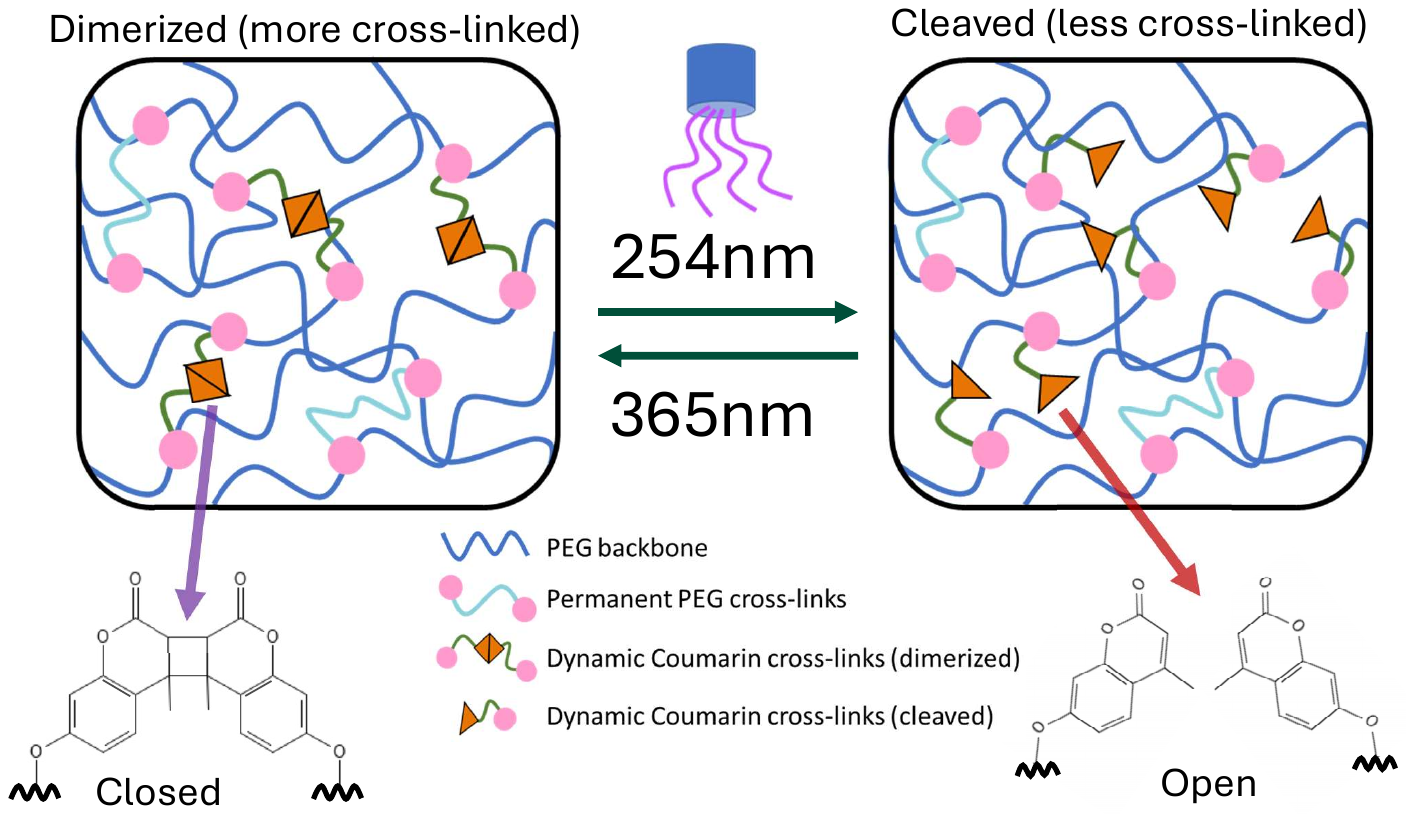}
    \caption{\textbf{Design of photoresponsive network.} Schematic depicting how this dual-cross-linked network is designed to switch between a more constrained state and a less constrained state upon post-cure irradiation with UV light.}
    \label{fig:network schematic}
\end{figure}

The CoumAc-functionalized PEG-based gel network developed here is designed to undergo bulk network transformations when exposed to UV light after polymerization (\textbf{Fig. \ref{fig:network schematic}}). 
Irradiation at 365 nm induces a [2+2] cycloaddition, forming cyclobutane-linked CoumAc dimers, whereas 254 nm irradiation cleaves these dimers, restoring CoumAc to its non-paired state.
The base PEG network, covalently cross-linked with PEGDA, maintains structural integrity regardless of post-cure irradiation. 
By incorporating CoumAc as a secondary, dynamic cross-linker, cross-link density can be reversibly increased \textit{via} 365 nm dimerization or decreased \textit{via} 254 nm cleavage.
This section outlines (i) how surface instabilities evolve during swelling as a function of CoumAc state (3.1), (ii) baseline kinetic, swelling, and mechanical properties of these networks (3.2), and (iii) chemical/mechanical changes during active photo-modulation (3.3).
Ultimately, the chemistry demonstrated here offers a route to design polymer surfaces with precisely programmable topographies \textit{via} remote, spatiotemporal control.
A critical prerequisite for such smart surfaces is a rigorous understanding of how CoumAc incorporation modulates both the bulk mechanical properties and the dynamic network transformations.

\subsection{Surface crease formation during gel swelling based on CoumAc state}
\begin{figure}[H]
    \centering
    \includegraphics[scale=0.6, angle=0, trim={1cm 6cm 1cm 6cm},clip]{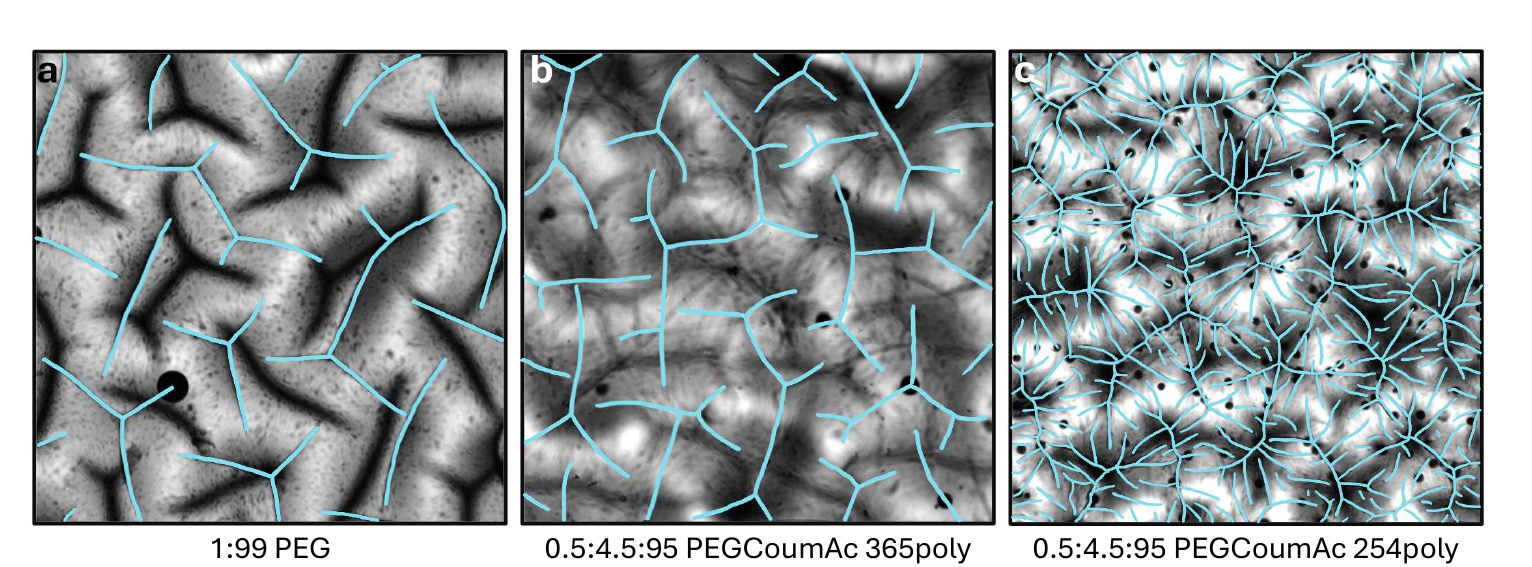}
    \caption{\textbf{The addition of CoumAc cross-linking, as well as the wavelength of UV light used during photo-polymerization, significantly changes crease patterns.} Microscope images measuring 3.328 x 3.328 mm compare the surface crease patterns on the surface of previously polymerized gels after 320 seconds of swelling in water for the \textbf{a)} 1:99 PEG formulation, as well as the 0.5:4.5:95 PEGCoumAc formulation \textbf{b)} polymerized at 365 nm and \textbf{c)} polymerized at 254 nm. In-focus creases are traced in blue to clarify the patterns of interest. Original microscope images are included in \textbf{Fig. S\ref{fig:micro images wo tracing}}.}
    \label{fig:microscope images}
\end{figure}

Polymer gels often develop surface instabilities such as creasing during swelling \cite{VanZanten2024, Louf2021, Leslie2021}. 
To asses how CoumAc cross-linking modulates these instabilities, we performed \textit{in situ} microscopy during swelling in water. 
Notably, all samples were fully polymerized before swelling, ensuring that the acrylate double bonds were consumed and that subsequent structural evolution could be attributed specifically to the dynamic CoumAc cross-links, rather than acrylate polymerization (representative conversion profiles and polymerization rate profiles are included in Supplementary information - \textbf{Figs. S\ref{fig:conv vs time}, S\ref{fig:Rp vs conv}}).
\textbf{Fig. \ref{fig:microscope images}} presents surface crease patterns for gels photopolymerized and then swollen in water for 320 seconds (here, swelling proceeds without any UV exposure). Three representative formulations are shown: 1:99 PEG (no CoumAc), 0.5:4.5:95 PEGCoumAc polymerized at 365 nm, and 0.5:4.5:95 PEGCoumAc polymerized at 254 nm. 
As a reminder, the formulation names indicate the mol\% of PEGDA, CoumAc, and PEGMA, respectively.
In the micrographs, thick, blurry lines correspond to out-of-focus creases on the top of the sample. 
Round dark features denote trapped air bubbles beneath the gel. 
The traced blue lines highlight in-focus creases on the bottom of the sample. 
(Unprocessed images are provided in \textbf{Fig. S\ref{fig:micro images wo tracing}}.)

Creases arise when the gel surface buckles under compressive stresses, leading to regions where the surface folds and contacts itself. 
These instabilities originate from gradients in swelling: 
as solvent is absorbed, the highly swollen outer layer is constrained by the initially unswollen core, producing depth-wise stress gradients.
Comparing the 1:99 PEG sample (\textbf{Fig. \ref{fig:microscope images}a}) with the 0.5:4.5:95 PEGCoumAc sample polymerized at 365 nm (\textbf{Fig. \ref{fig:microscope images}b}), we observe that the presence of CoumAc modestly increases the density of surface creases, although the general branched morphology - points where three creases meet - is preserved.
Notably, the crease density increase occurs despite similar swelling ratios at 300 seconds ($Q_{t}$ of 1.95 for 1:99 PEG, 1.71 for 0.5:4.5:95 365poly, and 1.68 for 0.5:4.5:95 254poly - see \textbf{Fig. S\ref{fig:sr data corr to images}}).
This indicates that differences in creasing morphology cannot be attributed solely to differences in swelling kinetics, but rather to changes in the network structure induced by the CoumAc cross-link and polymerization conditions.
In contrast, the same CoumAc formulation polymerized at 254 nm exhibited a markedly different brush-like pattern with very high crease density (\textbf{Fig.  \ref{fig:microscope images}c}) - implying greater compressive stresses, consistent with classical wrinkling theory linking stress magnitude to feature wavelength and density \cite{VanZanten2024}. This result is surprising given that gels polymerized at 254nm do not contain CoumAc dimers.
These differences underscore the profound influence of photopolymerization conditions and dynamic cross-linking on the evolution of surface instabilities.

\subsection{Network formation kinetics and swelling dynamics}
To assess how CoumAc cross-linking influences swelling-induced instabilities, we first established baseline chemical and mechanical properties directly following free-radical polymerization.
Real-time Fourier Transfer Infrared Spectroscopy (RT-FTIR) was used to quantify the effect of CoumAc content and irradiation wavelength during photopolymerization on network formation kinetics.
\textbf{Table \ref{tab:photopolymerization kinetics parameters}} reports the maximum polymerization rate ($R_P^{max}$) for each formulation cured at either 365 or 254 nm, along with the corresponding time to $R_P^{max}$ and final fractional conversion, calculated \textit{via} Equation \ref{fractional conversion}. 
A representative kinetic trace is provided in \textbf{Fig. S\ref{fig:conv vs time}}. 

Formulations containing CoumAc exhibit slower polymerization kinetics than the base PEG system, as evidence by markedly higher $R_P^{max}$ values for CoumAc-free formulations (highlighted in \colorbox{SkyBlue}{blue} in \textbf{Table \ref{tab:photopolymerization kinetics parameters}}).
Increasing CoumAc fraction progressively reduces $R_P^{max}$ and increases the time required to reach this maximum rate. 
This trend likely arises from the steric bulk of the CoumAc moiety, which can hinder favorable molecule orientation for rapid photopolymerization - particularly once the forming PEG network restricts segmental mobility.

\begin{table}[H]
    \centering
        \caption{\textbf{The rate of the free radical polymerization reaction is significantly faster at 365 nm than at 254 nm.} Photopolymerization kinetics parameters obtained \textit{via} RT-FTIR for all PEGCoumAc formulations for curing at both 365 and 254 nm, as well as for the base PEG system containing no CoumAc.}
    \begin{tabularx}{1\linewidth}{|c|c|c|c|c|c|}
    \cline{1-6}
        & & \multicolumn{4}{c|}{Photo-polymerized at 365nm}\\
        \cline{3-6}
        PEGDA & CoumAc & $R_p^{max}$ & Time at & Acrylate Fractional & Final Fractional\\
        (mol\%) & (mol\%) & (1/min.) & $R_p^{max}$ (min.) & Conversion at $R_p^{max}$ & Conversion\\
        \cline{1-6}
        0.5 & 0.5 & 1.24$\pm$0.47 & 0.33$\pm$0.21 & 0.22$\pm$0.10 & 1\\
        \cline{1-6}
        0.5 & 4.5 & 1.93$\pm$0.10 & 0.16$\pm$0.018 & 0.24$\pm$0.019 & 1\\
        \cline{1-6}
        \rowcolor{SkyBlue}
        1 & 0 & 4.44$\pm$0.06 & 0.11$\pm$0.005 & 0.25$\pm$0.001 & 1\\
        \cline{1-6}
        1 & 1 & 2.57$\pm$1.29 & 0.16$\pm$0.08 & 0.16$\pm$0.08 & 1\\
        \cline{1-6}
        1 & 4 & 1.85$\pm$0.55 & 0.15$\pm$0.02 & 0.22$\pm$0.04 & 1\\
        \cline{1-6}
        1 & 9 & 1.05$\pm$0.1 & 0.30$\pm$0.12 & 0.21$\pm$0.08 & 1\\
        \cline{1-6}
        \rowcolor{SkyBlue}
        5 & 0 & 4.87$\pm$0.06 & 0.12$\pm$0.004 & 0.29$\pm$0.004 & 1\\
        \cline{1-6}
        5 & 5 & 2.80$\pm$1.31 & 0.11$\pm$0.07 & 0.20$\pm$0.11 & 1\\
        \cline{1-6}
        5 & 15 & 0.85$\pm$0.05 & 0.53$\pm$0.01 & 0.33$\pm$0.01 & 1\\
        \cline{1-6}
        \rowcolor{SkyBlue}
        10 & 0 & 5.19$\pm$0.64 & 0.14$\pm$0.009 & 0.33$\pm$0.014 & 1\\
        \cline{1-6}
        & & \multicolumn{4}{|c|}{Photo-polymerized at 254nm}\\
        \cline{3-6}
        PEGDA & CoumAc & $R_p^{max}$ & Time at & Acrylate Fractional & Final Fractional\\
        (mol\%) & (mol\%) & (1/min.) & $R_p^{max}$ (min.) & Conversion at $R_p^{max}$ & Conversion\\
         \cline{1-6}
        0.5 & 0.5 & 0.60$\pm$0.13 & 0.72$\pm$0.64 & 0.14$\pm$0.09 & 0.97$\pm$0.04\\
        \cline{1-6}
        0.5 & 4.5 & 0.36$\pm$0.22 & 1.96$\pm$0.58 & 0.23$\pm$0.03 & 0.96$\pm$0.03\\
        \cline{1-6}
        1 & 1 & 0.44$\pm$0.04 & 1.51$\pm$0.42 & 0.28$\pm$0.12 & 0.98$\pm$0.0003\\
        \cline{1-6}
        1 & 4 & 0.24$\pm$0.16 & 5.05$\pm$4.27 & 0.40$\pm$0.02 & 0.98$\pm$0.01\\
        \cline{1-6}
        1 & 9 & 0.63$\pm$0.88 & 4.70$\pm$2.36 & 0.29$\pm$0.14 & 0.99$\pm$0.03\\
        \cline{1-6}
        5 & 5 & 0.28$\pm$0.08 & 3.72$\pm$0.97 & 0.41$\pm$0.01 & 0.99$\pm$0.01\\
        \cline{1-6}
        5 & 15 & 0.083$\pm$0.03 & 7.75$\pm$2.91 & 0.31$\pm$0.07 & 0.99$\pm$0.01\\
        \cline{1-6}
    \end{tabularx}
    \label{tab:photopolymerization kinetics parameters}
\end{table}

Polymerization kinetics differ substantially between the two cure wavelengths due to the optical properties of the photoinitiator DMPA, which absorbs roughly tenfold more strongly at 254 nm than at 365 nm \cite{absorbancespectra}. 
While higher absorbance generally accelerates radical generation, the effective rate is also governed by light intensity. 
Here, the 365 nm source delivered 0.1 $\frac{W}{cm^2}$ - over ten times the intensity of the 254 nm source (0.0088 $\frac{W}{cm^2}$). Furthermore, penetration depth of UV irradiation scales with wavelength (e.g., shorter wavelengths correspond to shorter penetration depths)\cite{LangInitiation2022}. The data summarized in \textbf{Table \ref{tab:photopolymerization kinetics parameters}} was collected on samples with thicknesses of $\sim$2 mm, which was chosen to enable capture and analysis of swelling-induced instabilities (\textbf{Fig. \ref{fig:microscope images}}).
Overall, these combined factors of intensity, absorption, and penetration depth resulted in faster curing at 365 nm despite DMPA's greater absorbance at 254 nm.
Accordingly, $R_p^{max}$ values at 365 nm were approximately an order of magnitude greater than those at 254nm (e.g., for the 0.5:0.5:99 formulation, $R_p^{max}=1.24$ at 365 nm, but $R_p^{max}=0.6$ at 254 nm). 
This trend was consistent across all formulations. 
Importantly, all samples achieved $\geq0.96$ final fractional conversion regardless of wavelength, indicating near-complete acrylate consumption.
Nonetheless, the slower network formation at 254 nm suggests significant differences in final network topology compared with the 365nm-cured gels. 

The fractional conversion at $R_p^{max}$ reflects the state of the network when auto-deceleration begins (see \textbf{Fig. S\ref{fig:Rp vs conv}} for representative $R_p$ vs. fractional conversion plots). 
Beyond $R_p^{max}$, chain mobility becomes increasingly restricted by rising viscosity and diffusion limitations imposed by the forming network \cite{Curley2024, Calvez2022}. 
For 365nm-cured samples, fractional conversions at $R_p^{max}$ were similar across formulations, whereas 254nm-cured samples displayed greater variation, consistent with the broader kinetic differences observed. Notably, at higher overall cross-link fraction, $R_p^{max}$ is shifted to higher degrees of conversion. 
Variations in the viscosity environment as a function of CoumAc fraction may also support the shift in $R_P^{max}$ to higher degrees of conversion. Modest increases in viscosity can enhance autoacceleration (e.g. diffusion-limited termination).

To contextualize these kinetic trends, the rate of initiation ($R_i$) was estimated using Eq. \ref{Ri}, which incorporates the initiation efficiency ($f$), quantum yield ($Q$), molar absorbtivity ($\epsilon$), concentration of initiator ($[I]$), light intensity ($I_0$), and wavelength ($\lambda$) \cite{odian}:
\begin{equation}\label{Ri}
    R_i =\frac{2fQ\epsilon[I]I_0\lambda}{0.1196 (J\cdot m/mol)}
\end{equation}

Using reported molar absorptivities for DMPA \cite{Lovell}, an initial efficiency $f=1$ (valid at early reaction stages \cite{LangInitiation2022}), and a quantum yield $Q=0.42$ \cite{Kowollik2017}, the ratio of $R_i$ at 365 nm versus 254 nm irradiation is calculated by Eq. \ref{Ri ratio}:
\begin{equation}\label{Ri ratio}
    \frac{R_{i,365}}{R_{i,254}}=\frac{\epsilon_{365} I_{0,365} \lambda_{365}}{\epsilon_{254} I_{0,254} \lambda_{254}}=\frac{(110)(0.1\frac{W}{cm^2})(365nm)}{(8830)(0.0088\frac{W}{cm^2})(254nm)}=0.20
\end{equation}
This calculation yields $\frac{R_{i,365}}{R_{i,254}}=0.2$, illustrating that DMPA's higher absorbance at 254 nm favors more rapid polymerizations, provided irradiation is not significantly attenuated within the sample.
As a result, these different irradiation and curing conditions are expected to produce networks with distinct microstructures \cite{Ahmadi2024} and potential gradients in network heterogeneity.

Eq. \ref{Ri} assumes a thin film geometry with limited reduction in UV intensity through the sample depth. However, our study focuses on bulk specimens with thicknesses of 2 mm, where attenuation of irradiation is significant. This attenuation is demonstrated by the intensity drop measurements provided in \textbf{Table S\ref{tab:intensitydrop}}. Thus, the calculated $R_i$ values cannot be correlated with our kinetic analyses (Table \ref{tab:photopolymerization kinetics parameters}). However, the ratio $\frac{R_{i,365}}{R_{i,254}}$ offers useful insight into how irradiation source characteristics influence network kinetics, it does not fully account for the reduction in effective initiation rate in thicker samples due to light attenuation.

\begin{figure}[H]
    \centering
    \includegraphics[scale=0.6, angle=0, trim={0cm 5cm 0cm 5cm},clip]{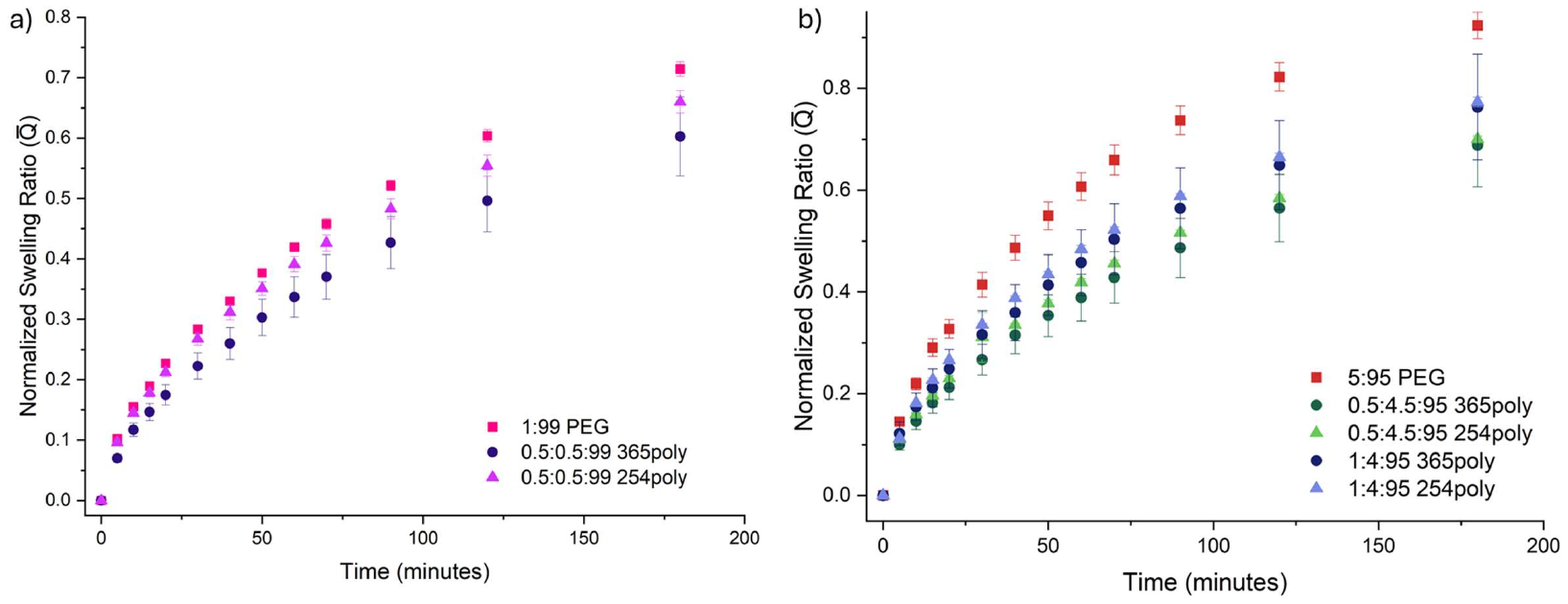}
    \caption{\textbf{Swelling behavior of PEGCoumAc samples polymerized at either 365nm or 254nm are compared to the base PEG system.} \textbf{a)} The normalized swelling ratio ($\overline{Q(t)}$) for PEGCoumAc formulations containing 0.5 mol\% CoumAc are compared to the 1:99 PEG formulation (polymerized at 365 nm). \textbf{b)} The normalized swelling ratio ($\overline{Q(t)}$) of PEGCoumAc formulations containing 0.5 and 1 mol\% CoumAc are compared to the 5:95 PEG formulation (polymerized at 365 nm).}
    \label{fig:qbar}
\end{figure}

To probe how these differing kinetic environments impact bulk measurements such as swelling behavior, 
we measured sample mass at discrete time points during immersion in water and calculated the normalized swelling ratio $\overline{Q(t)}$ (\textbf{Equation \ref{norm q}}).
\textbf{Fig. \ref{fig:qbar}} compares early-stage swelling (first 200 minutes) for formulations with identical total cross-linker content but varying PEGDA:CoumAc ratios and cure wavelengths.

\begin{figure}[H]
    \centering
    \includegraphics[scale=0.6, angle=0, trim={3cm 2.5cm 3cm 2.5cm},clip]{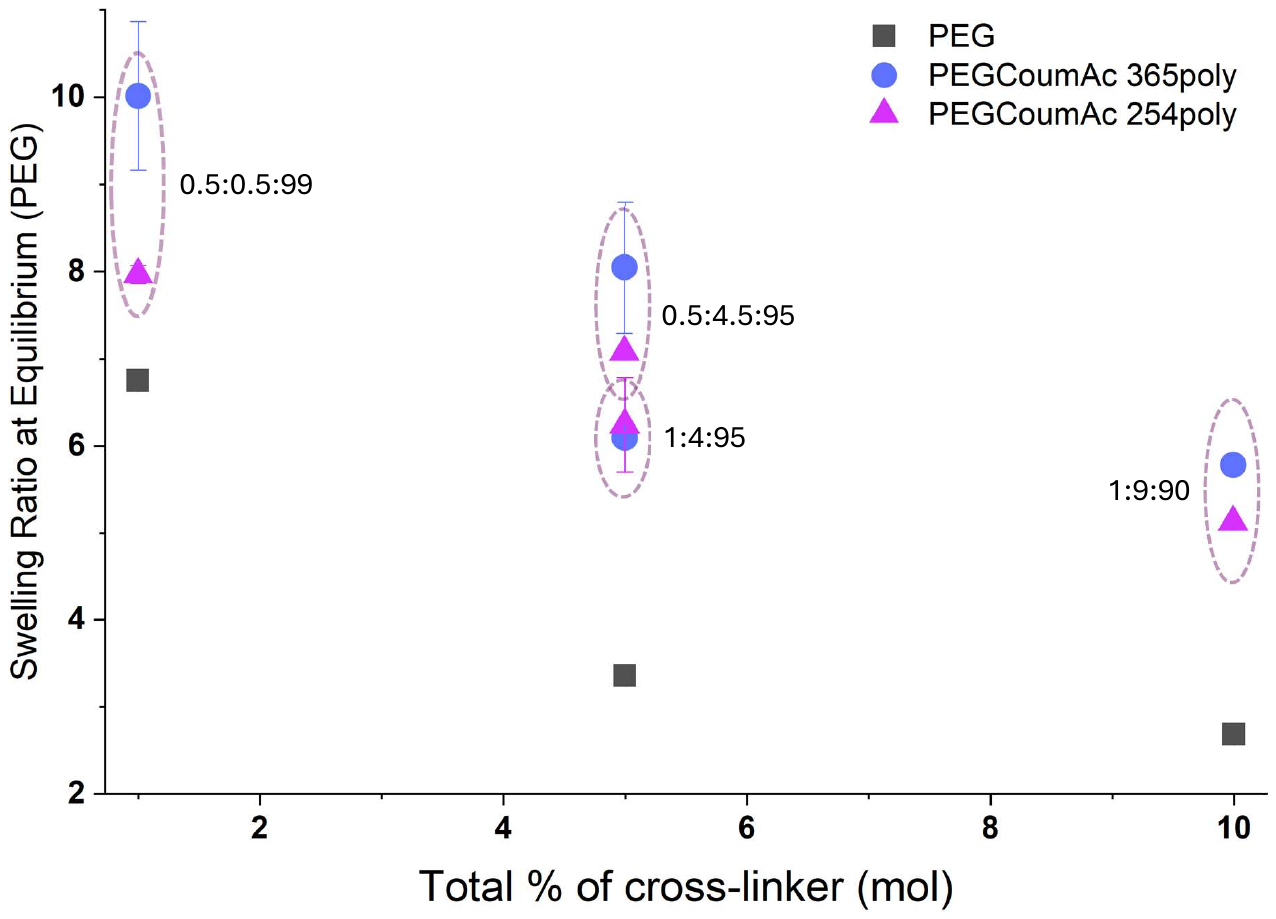}
    \caption{\textbf{Swelling ratio at equilibrium ($Q_{eq}$) is compared between PEG only formulations (polymerized at 365 nm), and PEGCoumAc formulations with equivalent total mol\% of cross-linker (polymerized at either 254 or 365 nm).}}
    \label{fig:qeq}
\end{figure}

For total cross-linker content of 1 mol\% (\textbf{Fig. \ref{fig:qbar}a}), replacing a portion of PEGDA with CoumAc reduced $\overline{Q(t)}$ relative to the PEGDA-only control (1:99 PEG) during early swelling, indicating slower approach to equilibrium.
At higher cross-linker fraction (5 mol\%, \textbf{Fig. \ref{fig:qbar}b}), a small increase in CoumAc fraction measurably decreased $\overline{Q(t)}$.
Furthermore, \textbf{Fig. \ref{fig:qeq}} shows that going from 4 mol\% CoumAc to 4.5 mol\% CoumAc increases the swelling ratio at equilibrium.
This increase in swelling capacity is likely due to the decrease in PEGDA fraction, and indicates that the reduction in $\overline{Q(t)}$ corresponds with longer time required to reach a higher equilibrium swelling ratio.
Across all cases, 254nm-cured samples showed marginally higher $\overline{Q(t)}$ in the first 200 minutes of swelling, though not always significant.
At equilibrium, swelling ratios $Q_{eq}$ (\textbf{Fig. \ref{fig:qeq}}) were higher for all CoumAc-containing gels than for PEGDA-only analogues of equal cross-link density, again reflecting CoumAc's lower cross-linking efficiency.

Interestingly, 
254nm-cured gels absorbed less water at equilibrium than 365nm-cured counterparts - opposite the expected trend if 365 nm curing produced more CoumAc dimers.
This reversal suggests that cure wavelength affects network topology in ways not captured by a simple cross-link count. Collectively, these results establish direct links between photopolymerization conditions, curing kinetics, and the resulting swelling properties of PEGCoumAc gels - providing the baseline context for interpreting photo-responsive changes that can be induced via \textit{post-cure} irradiation.

\subsection{Photo-responsive cross-linking kinetics: chemical and mechanical changes}
\begin{figure}[H]
    \centering
    \includegraphics[scale=0.55, angle=0, trim={0cm 5cm 0cm 5cm},clip]{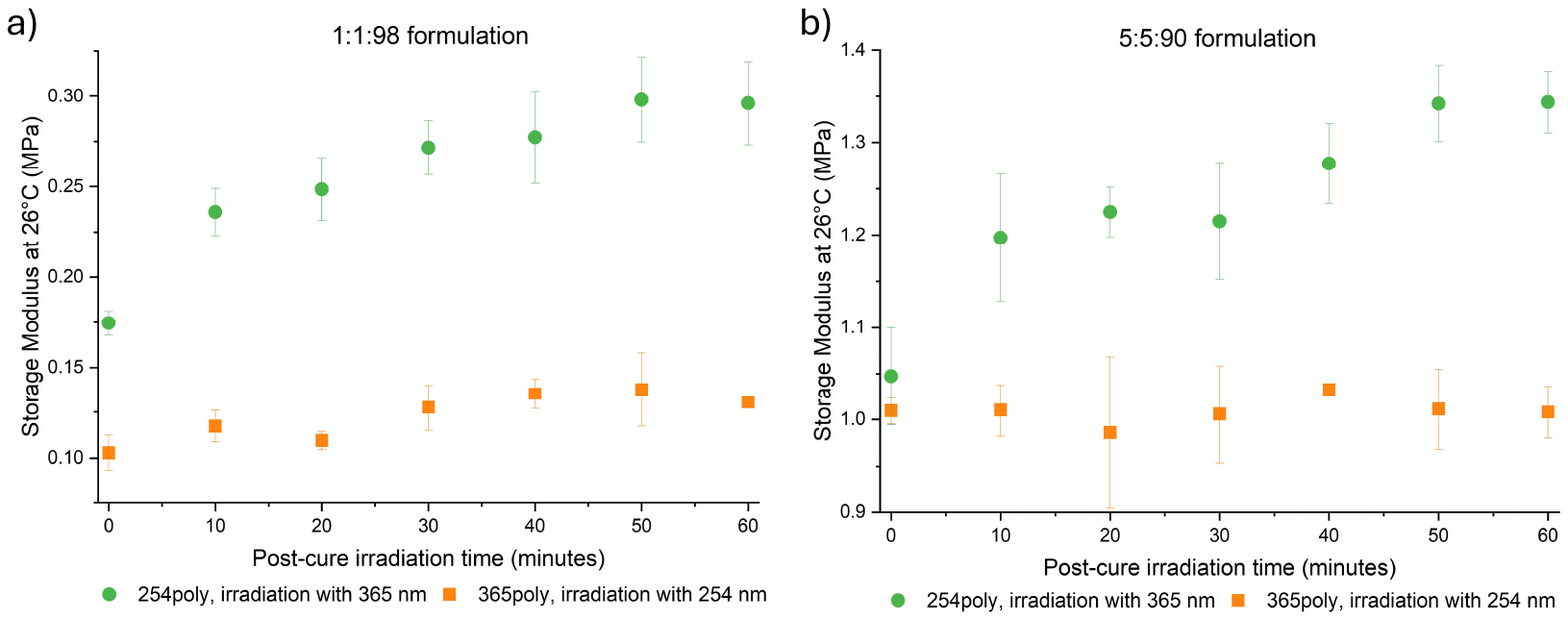}
    \caption{\textbf{Post-cure irradiation with 365 nm significantly increases storage modulus at room temperature, but no change is observed during irradiation with 254 nm.} The storage modulus at 26\degree C is measured at 10 minute intervals during post-cure irradiation for a) 1:1:98 samples and b) 5:5:90 samples. Samples polymerized at 365 nm are subsequently irradiated with 254 nm light (orange circle) to induce cleavage, while samples polymerized at 254 nm are irradiation with 365 nm light (green squares) to induce dimerization.}
    \label{SM vs irradiation time}
\end{figure}

Initial, static mechanical properties, measured \textit{via} DMA at 26\degree C (reported in \textbf{Fig. \ref{SM vs irradiation time}} at post-cure irradiation time equal to zero) revealed higher storage moduli for gels polymerized via 254 nm irradiation than for analogous counterparts polymerized at 365 nm - contrary to expectations that dimerization during 365 nm curing would stiffen the network.
This paradox can be rationalized by differences in polymerization kinetics.
The slower network growth at 254 nm allows polymer chains and CoumAc moieties more time to explore conformational space before gelation, promoting a more homogeneous cross-link distribution. 
Such structural uniformity enhances load-bearing efficiency, producing higher macroscopic stiffness.
By contrast, rapid curing at 365 nm may kinetically trap a less ordered, more heterogeneous network, lowering storage modulus despite higher potential cross-link density from CoumAc dimerization.

Post-cure irradiation triggering either CoumAc dimerization or cleavage is anticipated to alter both the molecular-scale and bulk mechanical properties as a function of irradiation time.
To quantify macroscopic effects, we monitored storage modulus ($E^{'}$) at 26\degree C \textit{via} DMA during post-cure UV exposure.
\textbf{Fig. \ref{SM vs irradiation time}} compares two reciprocal experiments: 
1) Cleavage: 254 nm irradiation of gels polymerized at 365 nm (orange series, squares), and
2) Dimerization: 365 nm irradiation of gels polymerized at 254 nm (green series, circles).
Data are shown for 1:1:98 gels (\textbf{Fig. \ref{SM vs irradiation time}a}) and 5:5:90 gels (\textbf{Fig. \ref{SM vs irradiation time}b}). 
In the cleavage case, storage modulus is essentially unchanged, implying that 7 minutes of 365 nm curing did not generate a substantial fraction of CoumAc dimers;
Thus, subsequent irradiation with 254 nm had minimal impact on bulk stiffness.
The observed attenuation of 254 nm UV light through the sample depth likely also limited any coumarin cleavage to the surface of the sample.

In contrast, post-cure dimerization \textit{via} 365 nm irradiation produced substantial stiffening.
After 60 minutes, $E^{'}$ increased by 69\% in the 1:1:98 formulation and by 28\% in the 5:5:90 formulation.
Strikingly, $\sim50\%$ of this total increase occurred with the first 10 minutes (35\% and 14\%, respectively) highlighting the rapid onset of network reinforcement once dimerization begins.
These results indicate that the CoumAc moieties embedded within the network retain sufficient mobility to undergo anomalous diffusion, enabling reactive encounters and [2+2] cycloadditions. 
Even a modest loading (1 mol\% of CoumAc) can induce a pronounced macroscopic response, with $E^{'}$ rising by $\geq65\%$ - demonstrating that low concentrations of dynamic cross-linker can effect large changes in bulk stiffness.

To probe chemical changes directly, we used pseudo-real-time mid-IR FTIR on 5:5:90 PEGCoumAc films of $\sim$0.01 mm thickness (to minimize absorbance saturation). In all experiments, a single sample was evaluated to minimize differences in functional group concentration as a function of variations in sample preparation.
Two diagnostic features of the coumarin chromophore were tracked: 
1) the aryl conjugated carbon-carbon double bond stretch ($\approx$1616 $cm^{-1}$), and 
2) the carbonyl (C=O) stretch ($\approx$1740 $cm^{-1}$) \cite{Inacker2022}. 
Upon dimerization, the C-C double bond signal is expected to diminish and shift slightly upward in wavenumber, while the carbonyl peak shifts to higher wavenumbers due to altered conjugation.

\begin{figure}[H]
    \centering
    \includegraphics[scale=0.55, angle=0, trim={0cm 5cm 0cm 4cm},clip]{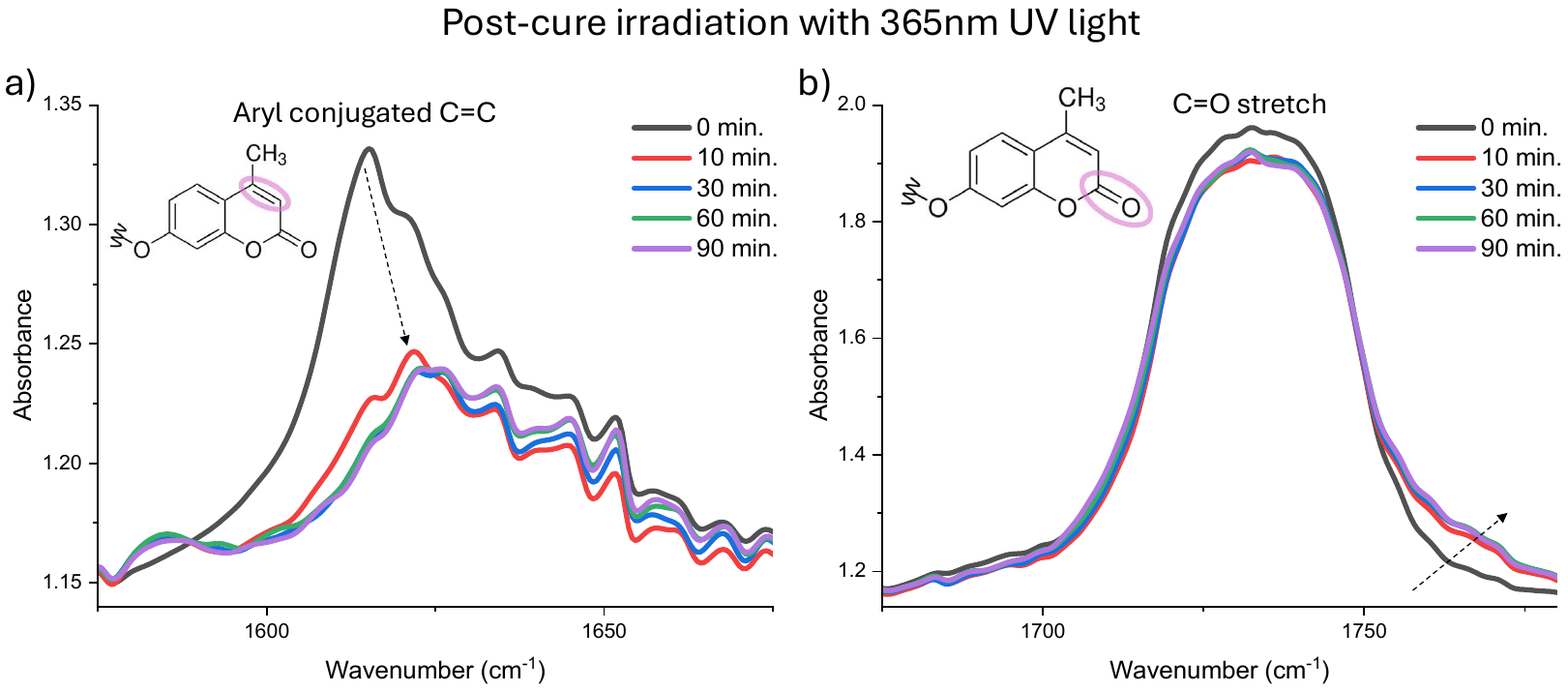}
    \caption{\textbf{A 5:5:90 PEGCoumAc film (0.01 mm thick) polymerized with 254 nm UV light for 2 minutes underwent subsequent 90 minutes of post-cure irradiaiton with 365 nm to induce dimerization of CoumAc molecules.} \textbf{a)} The peak corresponding to the C-C double bond present in the non-bonded CoumAc molecule is highlighted, and \textbf{b)} features the peak corresponding to the C=O stretching signal.}
    \label{fig:midir365}
\end{figure}

As shown in \textbf{Fig. \ref{fig:midir365}a}, the carbon-carbon double bond peak centered at 1616 $cm^{-1}$ decreased markedly in intensity and shifted to $\approx$1624 $cm^{-1}$ during 365 nm irradiation. Concurrently, the C=O peak shifted modestly toward higher wavenumbers (\textbf{Fig. \ref{fig:midir365}b}). 
Most spectral evolution occurred within the first 60 minutes, with negligible change thereafter - consistent with the stiffening kinetics observed by DMA (\textbf{Fig. \ref{fig:storage modulus at room temp}}).

\begin{figure}[H]
    \centering
    \includegraphics[scale=0.55, angle=0, trim={0cm 5cm 0cm 4cm},clip]{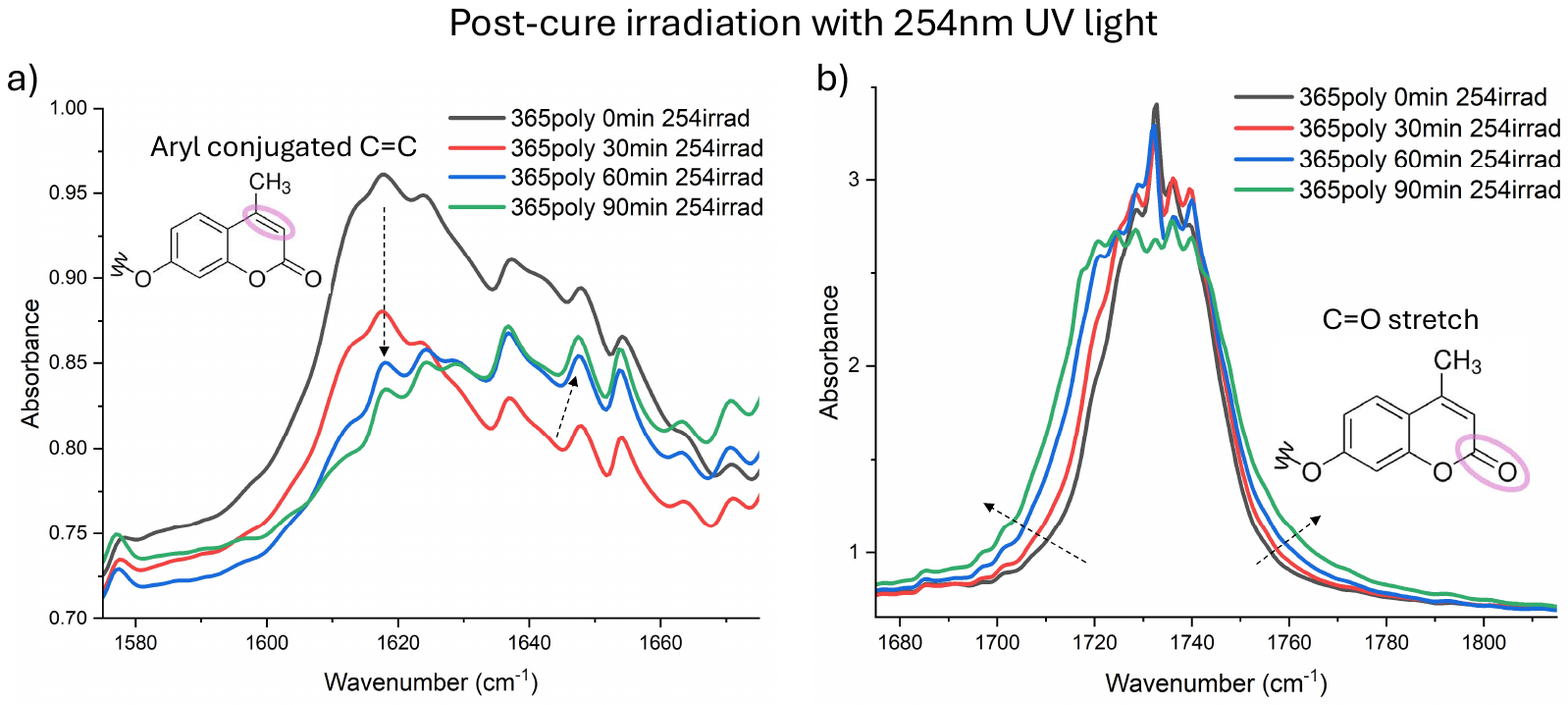}
    \caption {\textbf{A 5:5:90 PEGCoumAc film (0.01 mm thick) polymerized with 365 nm UV light for 90 minutes underwent subsequent 90 minutes of post-cure irradiaiton with 254 nm to induce cleavage of CoumAc molecules.} \textbf{a)} The peak corresponding to the C-C double bond present in the non-bonded CoumAc molecule is highlighted, and \textbf{b)} features the peak corresponding to the C=O stretching signal.}
    \label{fig:new254irrad}
\end{figure}

For cleavage kinetics, films were polymerized at 365 nm (90 minutes) to maximize dimer content, then exposed to 254 nm irradiation.
Initially, the 1616 $cm^{-1}$ C-C double bond peak decreased, but by 60 minutes a new shoulder emerged between 1637-1660 $cm^{-1}$ (see \textbf{Fig. \ref{fig:new254irrad}a}), likely reflecting vibrational modes of carbon centers adjacent to the carbonyl in the cleaved state - supporting dimer scission.
In parallel, the C=O band broadened (\textbf{Fig. \ref{fig:new254irrad}b}), also consistent with structural reorganization.

FTIR data indicate that both the dimerization and cleavage in the bulk network require $\geq60$ minutes to approach steady state. 
The unexpectedly slow cleavage is attributed to low 254 nm intensity and limited penetration depth into the millimeter scale network (intensity measurements provided in \textbf{Table S\ref{tab:intensitydrop}}). 
Previous investigations by Kabb et al. into cleavage kinetics \textit{via} UV-Vis report that 5-15 minutes of 254nm irradiation is required to effect significant increase of the absorbance of the C-C double bond reformation \cite{Kabb2018}.
By incorporating CoumAc into this PEG gel network, topological constraints are effectively increased, meaning that we'd expect cleavage in our system to be slower than that reported via UV-vis analysis.
Nonetheless, 10 minutes of 365 nm exposure suffices to drive substantial dimerization for CoumAc moieties already in close proximity (\textbf{Fig. \ref{SM vs irradiation time}}); more distant ones require longer times, with overall rates depending on total cross-link density and CoumAc loading.

\begin{figure}[H]
    \centering
    \includegraphics[width=1\linewidth]{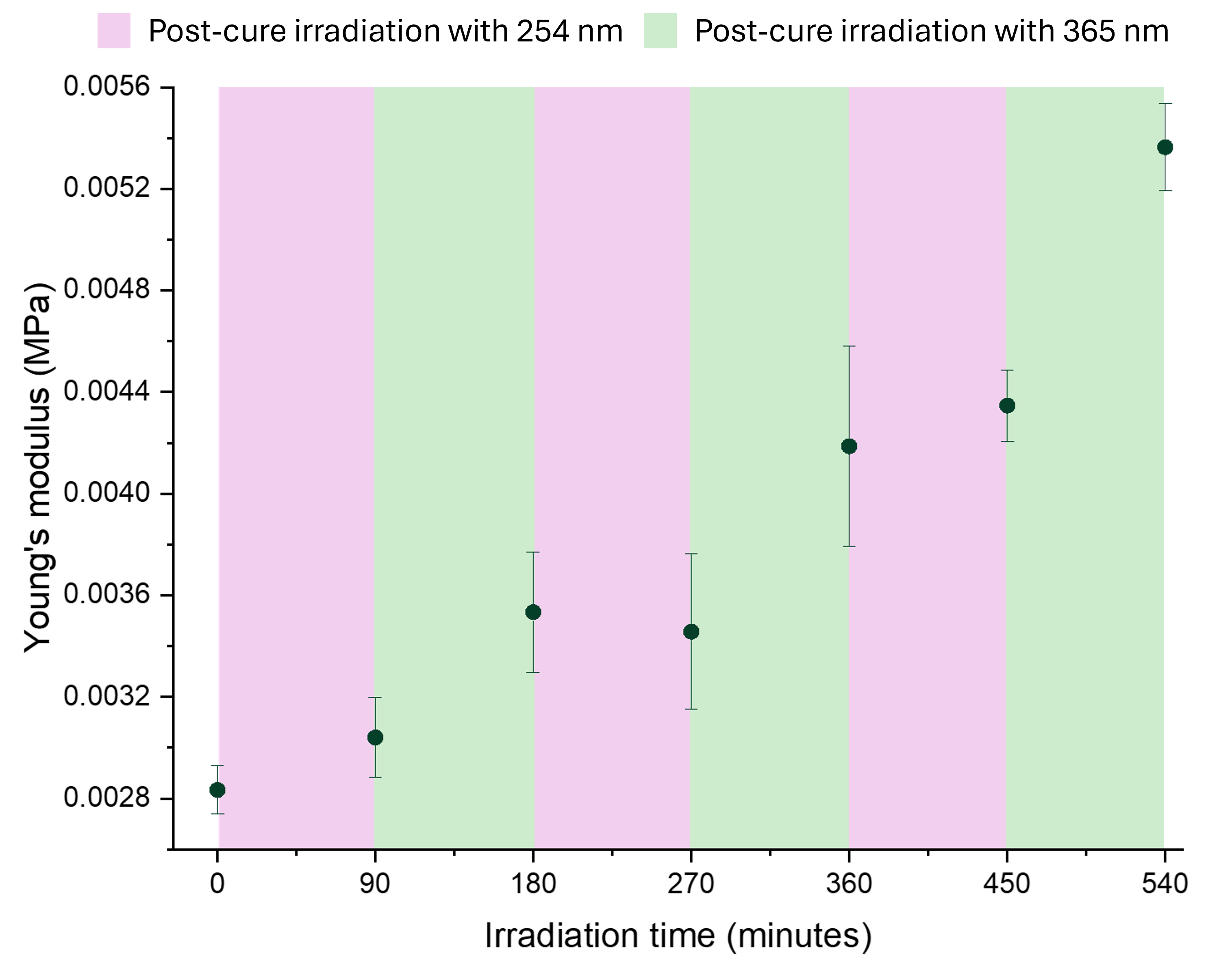}
    \caption{\textbf{Cycling between 254 nm and 365 nm irradiation revealed that cleavage has no significant impact on bulk mechanical properties.} A 1:9:90 sample polymerized at 365 nm for 77 minutes was irradiated, post-cure, with 254 nm UV light for 90 minutes (pink regions) and then with 365 nm for 90 minutes (green regions). This procedure was repeated for 3 total cycles.}
    \label{fig:cycling}
\end{figure}

To assess reversibility, 1:9:90 bars (25 \textit{x} 10 \textit{x} 2 mm) polymerized at 365 nm for 77 minutes were subjected to three cleavage/dimerization cycles: (i) 254 nm (90 minutes irradiation) (ii) 365 nm (90 minutes irradiation), which was repeated.
Young's modulus was measured after each step to track bulk property evolution (\textbf{Fig. \ref{fig:cycling}}).

With this experiment, 254 nm irradiation steps had negligible effect on Young's modulus, despite the expectation of modulus loss from dimer cleavage.
Combined with FTIR evidence (\textbf{Fig. \ref{fig:new254irrad}}), this suggests that cleavage at 254 nm is largely confined to the surface, given the attenuation through the sample depth, meaning that insufficient photons are delivered to measurably alter bulk load-bearing structure.
Conversely, each 365 nm irradiation step increased modulus further;
After 270 minutes cumulative exposure, modulus had doubled relative to post-cure baseline - implying that many CoumAc groups remained available for dimerization well after initial cure.
Irradiation at 365 nm on the scale of tens of hours would likely be required to reach full dimerization in the network.
The slower bulk kinetics compared with thin-film FTIR curves (\textbf{Fig. \ref{fig:midir365}}) highlight that network-wide dimerization is diffusion limited in thick specimens.
Overall, these results show that while efficient bulk stiffening \textit{via} dimerization is achievable on hour-scale timescales, bulk softening via cleavage is minimal under current 254 nm conditions.

\begin{figure}[H]
    \centering
    \includegraphics[scale=1, angle=0, trim={0cm 3cm 14cm 3cm},clip]{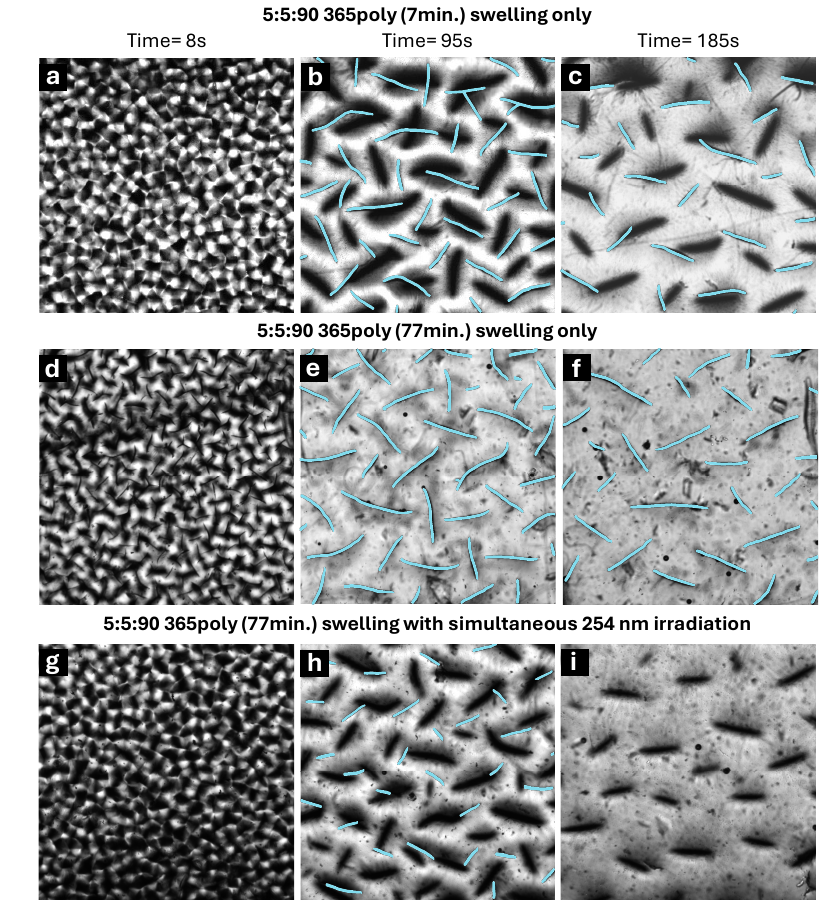}
    \caption{\textbf{\textit{In situ} microscopy reveals that irradiation with 254 nm UV light during swelling increases the rate of crease evolution.} Three 5:5:90 PEGCoumAc samples with varying polymerization conditions and post-cure exposure are shown at \textbf{a,d,g)} 8, \textbf{b,e,h)} 95, and \textbf{c,f,i)} 185 seconds of swelling. \textbf{a-c)} feature a sample polymerized at 365 nm for 7 minutes, \textbf{d-f)} show a sample polymerized at 365 nm for 77 minutes, and \textbf{g-i)} show the impact of simultaneous 254 nm irradiation during swelling for a sample polymerized at 365 nm for 77 minutes. All images show a field of view 3.328 x 3.328 mm.}
    \label{fig:254_irrad_impact_on_creases}
\end{figure}

To probe the impact of polymerization time and post-cure UV exposure on crease evolution, \textit{in situ} microscope images were collected during swelling for varying 5:5:90 PEGCoumAc samples polymerized at 365 nm (\textbf{Fig. \ref{fig:254_irrad_impact_on_creases}}). We first compare crease evolution during swelling between a  sample photopolymerized at 365 nm for a short irradiation period (7 minutes - \textbf{Fig. \ref{fig:254_irrad_impact_on_creases} a-c}) and a longer irradiation period (77 minutes - \textbf{Fig. \ref{fig:254_irrad_impact_on_creases} d-f}). These two curing protocols were informed by the kinetic data in \textbf{Fig. S\ref{fig:conv vs time}} and the storage modulus data in \textbf{Fig. S\ref{fig:storage modulus at room temp}}. Our kinetic profiles highlight that a 7 minute curing period is sufficient for complete double bond consumption (i.e., 1.0 fractional conversion of acrylate functionalities), while our DMA data informs that the storage modulus plateaued after 70 minutes of additional 365 nm irradiation.
Comparing these two sets of images reveal qualitative similarities between the transient crease evolution: dense crease formation at early time points (first few seconds of swelling) as well as similar morphologies (i.e. lack of branching) over the following minutes. 
This comparison also reveals that prolonged 365 nm exposure time for photopolymerization (77 minutes), which increases dimer content, results in creases that are both longer and more densely packed at 185 seconds of swelling. This suggests that higher dimer content suppresses swelling kinetics and maintains elevated compressive stresses for a longer duration during swelling.

Given the mid-FTIR results thinner specimens of this formulation (\textbf{Figs. \ref{fig:new254irrad}}), we hypothesize that at a minimum, photo-mediated events will arise at the surface of our gel samples, which may alter dynamic crease evolution during swelling if a sample is exposed to 254 nm irradiation \textit{during} swelling (\textbf{Fig. \ref{fig:254_irrad_impact_on_creases} g-i}). Our images reveal that this is the case, when a sample that is cured via 365 nm irradiation for 77 minutes and thus has a high dimer content, is exposed to 254 nm irradition during swelling, thus promoting cleavage of CoumAc, surface creases rapidly disappear, with in-focus creases absent at 185 second (\textbf{Fig. \ref{fig:254_irrad_impact_on_creases} i}).
This observation indicates that dimer cleavage by 254 nm light accelerates swelling and relieves surface compressive stresses during the initial swelling period.

Together, these results demonstrate that post-cure photomodulation - especially, \textit{in situ} 254 nm irradiation - provides an effective means of controlling surface creasing through dynamic manipulation of the network's cross-linking state during swelling.
The ability to tune crease morphology in real time by moving from dimerized to cleaved CoumAc cross-linking highlights the potential for spatiotemporal programming of surface patterns in photoresponsive polymer gels.
Therefore, this PEGCoumAc gel is considered a worthwhile candidate for future investigations into engineered control of interfacial behavior.

\section*{Conclusions}
This study demonstrates that coumarin-functionalized PEG-gels provide a versatile platform for dynamically tuning mechanical properties, swelling behavior, and surface topography through controlled UV irradiation.
By integrating permanent (PEGDA) and photo-responsive (CoumAc) cross-linkers within a single  network, we modulate cross-link density post-cure \textit{via} light-driven coumarin dimerization and cleavage.
Surface crease imaging revealed CoumAc incorporation and photopolymerization conditions (wavelength, intensity, duration) strongly influence instability morphology, density, and evolution during swelling.
These findings indicate that dynamic cross-links alter internal stress-relaxation pathways and can encode distinctive topographic patterns.
Furthermore, irradiation of a 365nm-cured gel \textit{in situ} with 254 nm UV light was shown to accelerate network relaxation and surface creases disappeared more quickly.

Swelling experiments show that CoumAc is a less efficient swelling-resistant cross-linker than PEGDA, resulting in higher equilibrium swelling ratios when PEGDA is replaced.
Nevertheless, post-cure 365 nm dimerization significantly increases stiffness - by up to 69\% - on hour or sub-hour timescales, demonstrating that small CoumAc loadings can produce substantial, rapid mechanical reinforcement.
Real-time FTIR confirmed that both dimerization and cleavage occur in the bulk network, though kinetics are sensitive to irradiation protocol and initial network structure. 
Future work aims to further explore this spectroscopic data to quantify the extent of the dimerization and cleavage reactions.

Importantly, DMA cycling experiments revealed that bulk softening via 254 nm cleavage is minimal under current irradiation conditions, likely due to 254 nm attenuation through the sample, suggesting that practical property modulation is surface-confined for cleavage but bulk-accessible for dimerization.
Accounting for UV attenuation in bulk samples is key to the future design of photoresponsive hydrogels, as attenuation can limit the depth and degree of cross-linker activation or cleavage, especially for wavelengths with high absorbance but low penetration (254 nm).

Overall, the interplay between photochemisty, mechanical response, and swelling induced stress patterns offers a powerful materials-design lever.
The dual-cross-linked PEGCoumAc framework enables spatiotemporal programming of stiffness, internal stress profiles, and surface morphology - capabilities highly relevant to adaptive coatings, reconfigurable membranes, soft actuators, and responsive biomaterials.
Future work will optimize irradiation parameters to improve cleavage efficiency and quantify surface-localized transformation.

\section*{Acknowledgments}
During this study, Alyssa VanZanten and Surbhi Punhani-Schillinger were supported by the NSF/ DMR (Award 2311697); Shih-Yuan Chen was partially supported by the NSF/DMR (Award 2311698). 

\newpage
\singlespacing
\bibliographystyle{unsrt}
\bibliography{bibliography}
\newpage
\section*{Supplementary Information}
\setcounter{figure}{0}

\begin{figure}[H]
    \centering
    \includegraphics[width=1\linewidth]{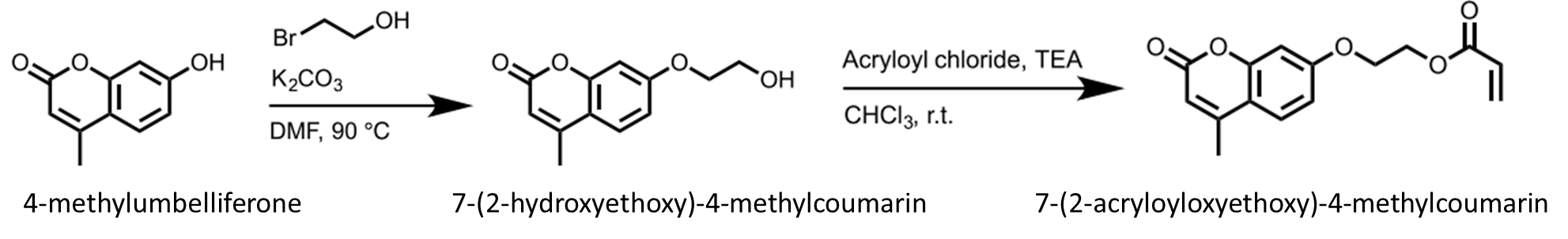}
    \caption{\textbf{Reaction scheme for the synthesis of 7,2-(acryloyloxyethoxy)-4-methylcoumarin.}}
    \label{fig:rxn scheme}
\end{figure}

\begin{figure}[H]
    \centering
    \includegraphics[width=1\linewidth]{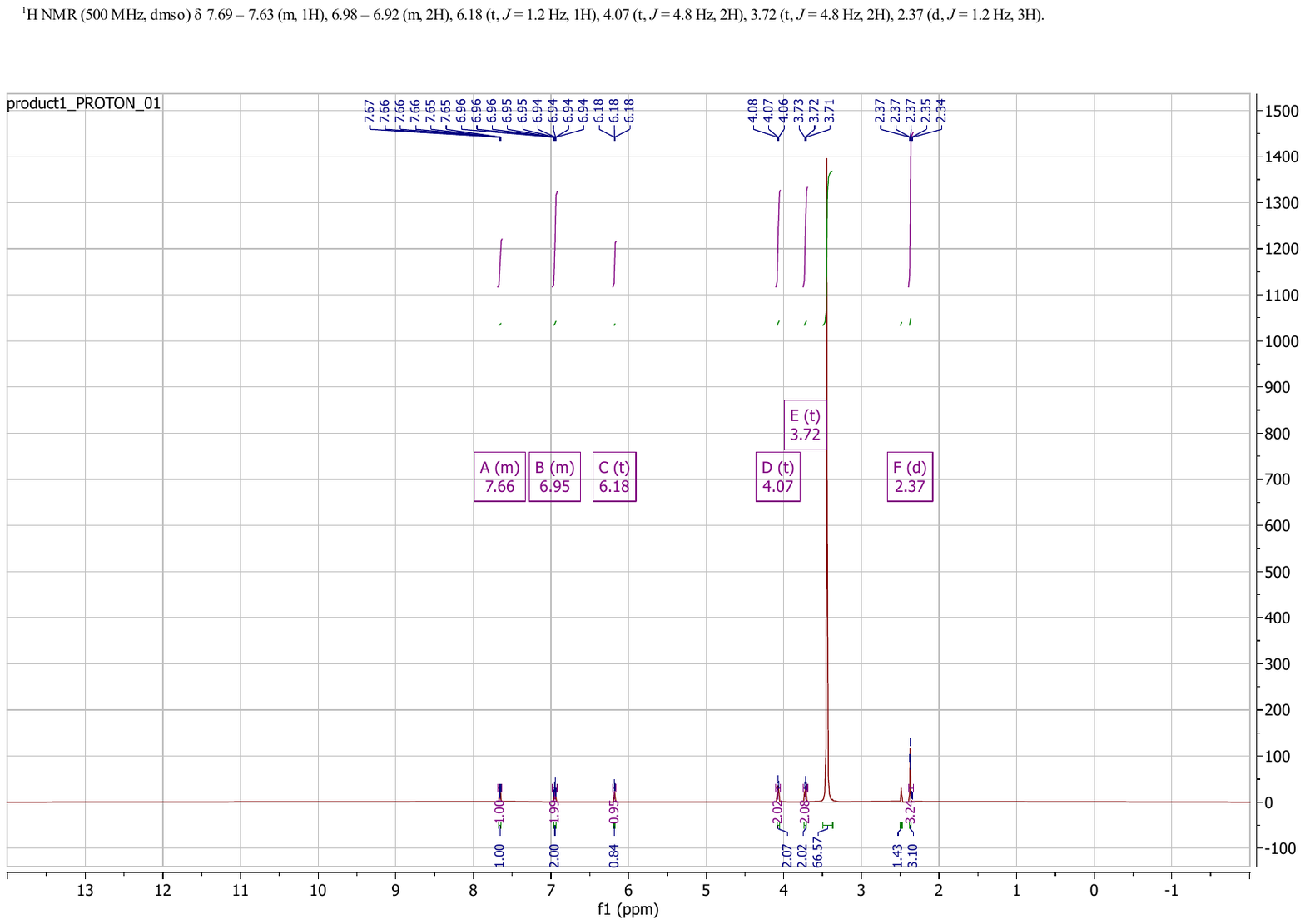}
    \caption{H-NMR spectrum of intermediate product of synthesis}
    \label{fig:prod1 nmr}
\end{figure}

\begin{figure}[H]
    \centering
    \includegraphics[width=1\linewidth]{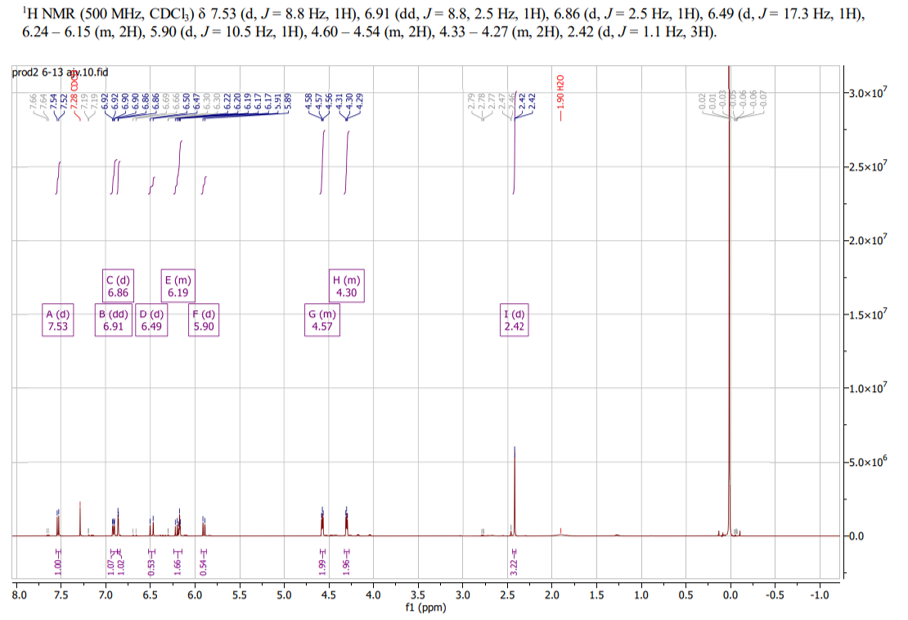}
    \caption{H-NMR spectrum of final synthesis product}
    \label{fig:prod2 nmr}
\end{figure}

\begin{figure}[H]
    \centering
    \includegraphics[width=1\linewidth]{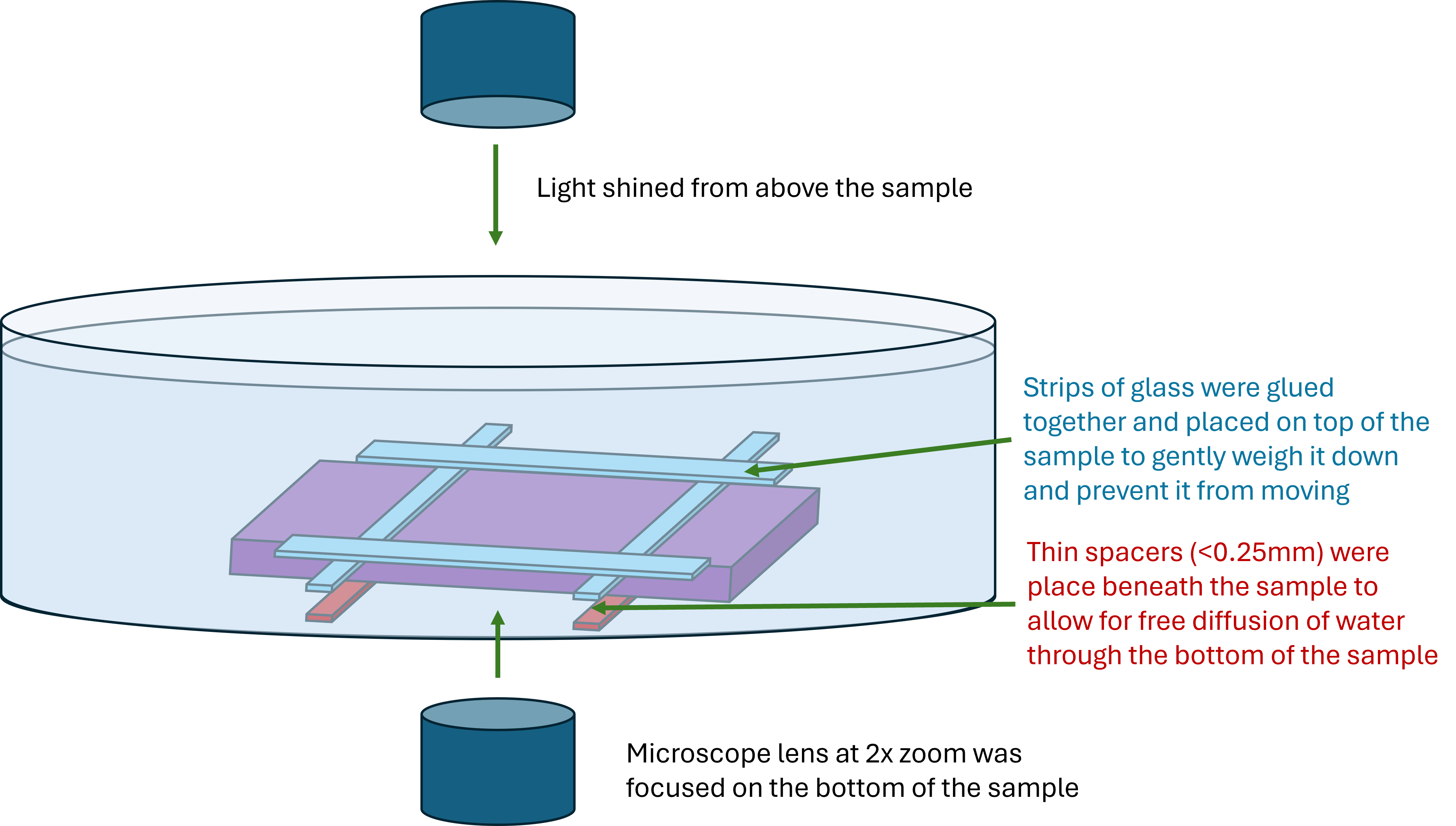}
    \caption{\textbf{A description of the setup used to ensure swelling samples did not move around too much during microscope imaging.}}
    \label{fig:microscope setup}
\end{figure}

\begin{figure}[H]
    \centering
    \includegraphics[width=1\linewidth]{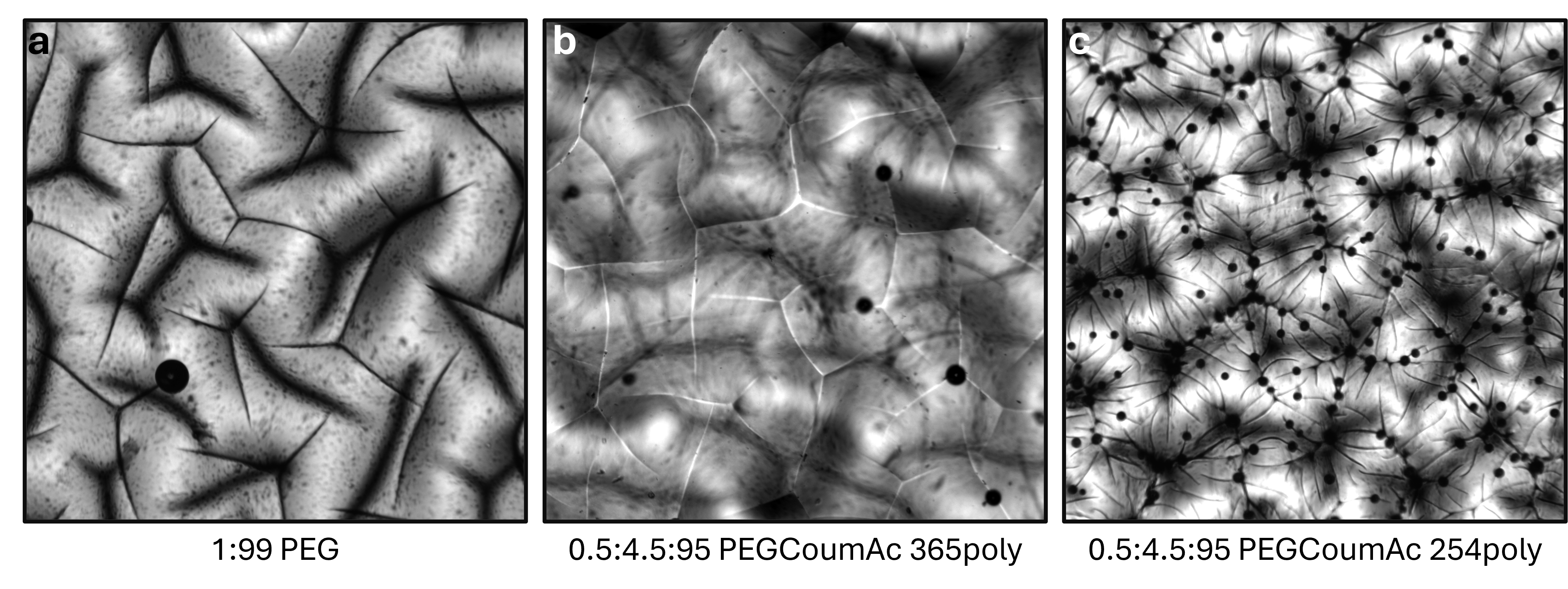}
    \caption{\textbf{Original microscope images without tracing} for \textbf{a)} a 1:99 PEG sample, \textbf{b)} a 0.5:4.5:95 PEGCoumAc sample polymerized at 365 nm, and \textbf{c)} a 0.5:4.5:95 PEGCoumAc sample polymerized at 254 nm. All images are at 320 seconds of swelling and show a field of view 3.328 x 3.328 mm. }
    \label{fig:micro images wo tracing}
\end{figure}

\begin{figure}[H]
    \centering
    \includegraphics[width=1\linewidth]{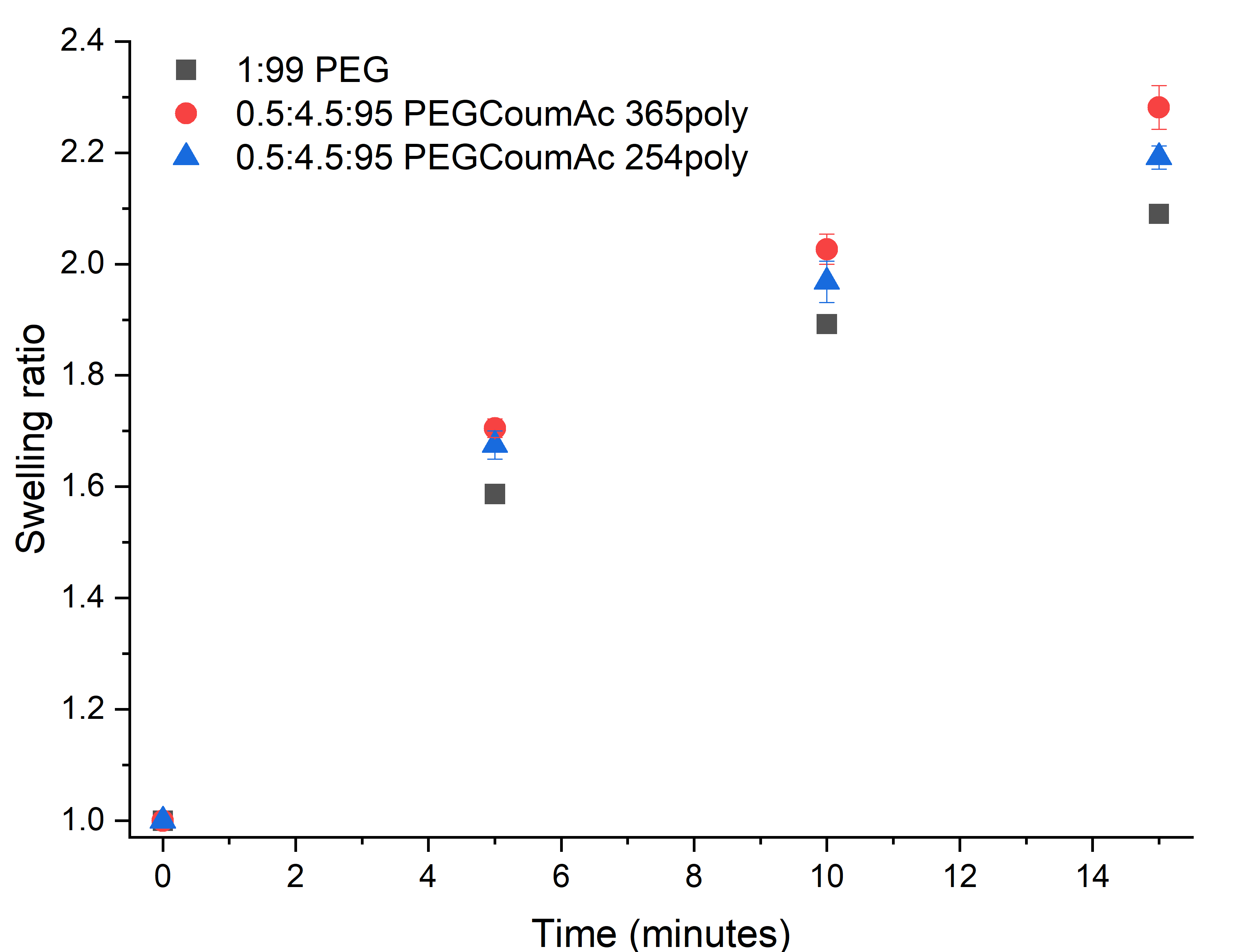}
    \caption{\textbf{Swelling ratio ($Q$) during the first 15 minutes of swelling is included for the 1:99 PEG formulation as well as the 0.5:4.5:95 PEGCoumAc formulation under either 365 nm or 254 nm polymerization conditions.}}
    \label{fig:sr data corr to images}
\end{figure}

\begin{figure}[H]
    \centering
    \includegraphics[width=0.5\linewidth]{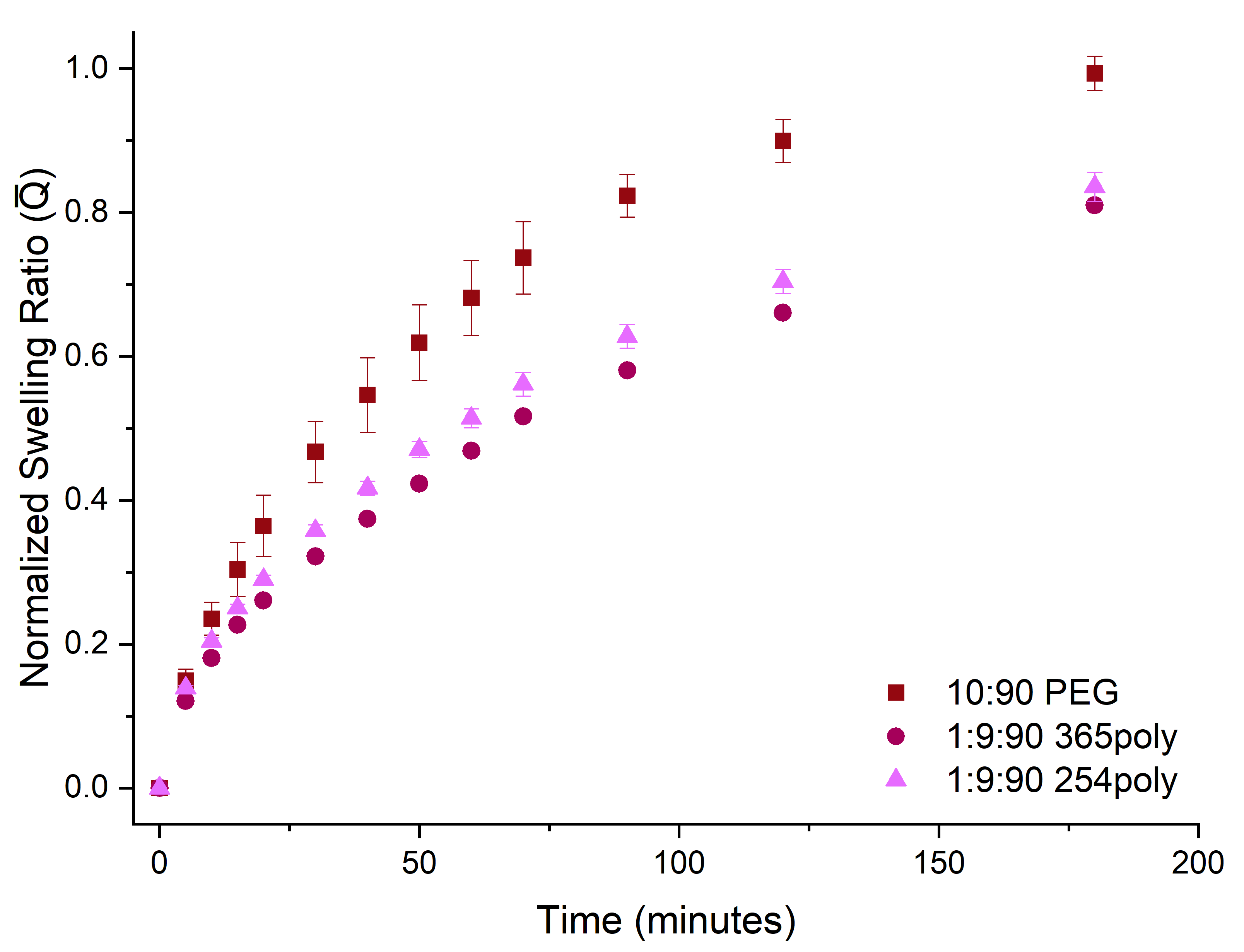}
    \caption{\textbf{Normalized swelling ratio ($\overline{Q}$) for the 1:9:90 PEGCoumAc formulation polymerized at either 365 nm or 254 nm are compared to the 10:90 PEG formulation over the first 200 minutes of swelling.}}
    \label{fig:10totalmol}
\end{figure}

\begin{figure}[H]
    \centering
    \includegraphics[width=1\linewidth]{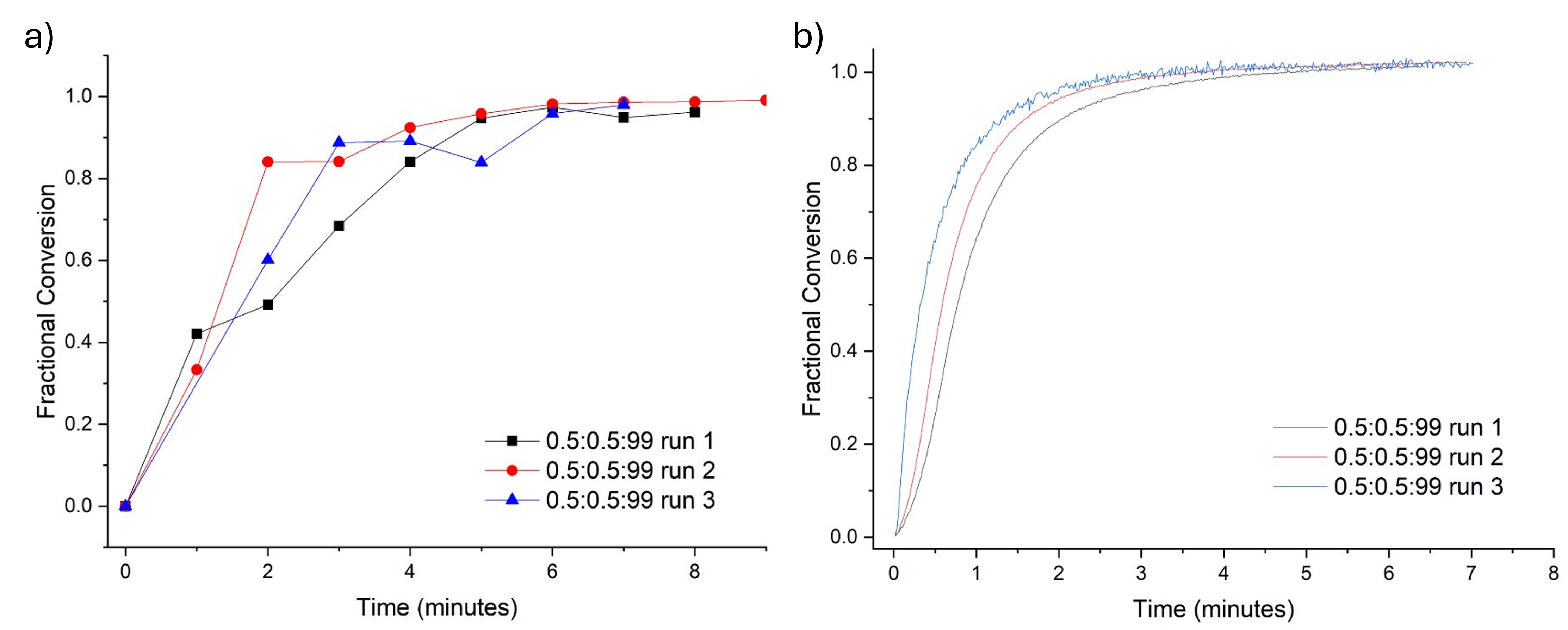}
    \caption{\textbf{Examples of the polymerization curves collected during curing.} \textbf{a)} The pseudo-real-time 254 nm and \textbf{b)} real-time 365 nm polymerization kinetics are compared for the 0.5:0.5:99 PEGCoumAc formulation.}
    \label{fig:conv vs time}
\end{figure}

\begin{figure}[H]
    \centering
    \includegraphics[width=1\linewidth]{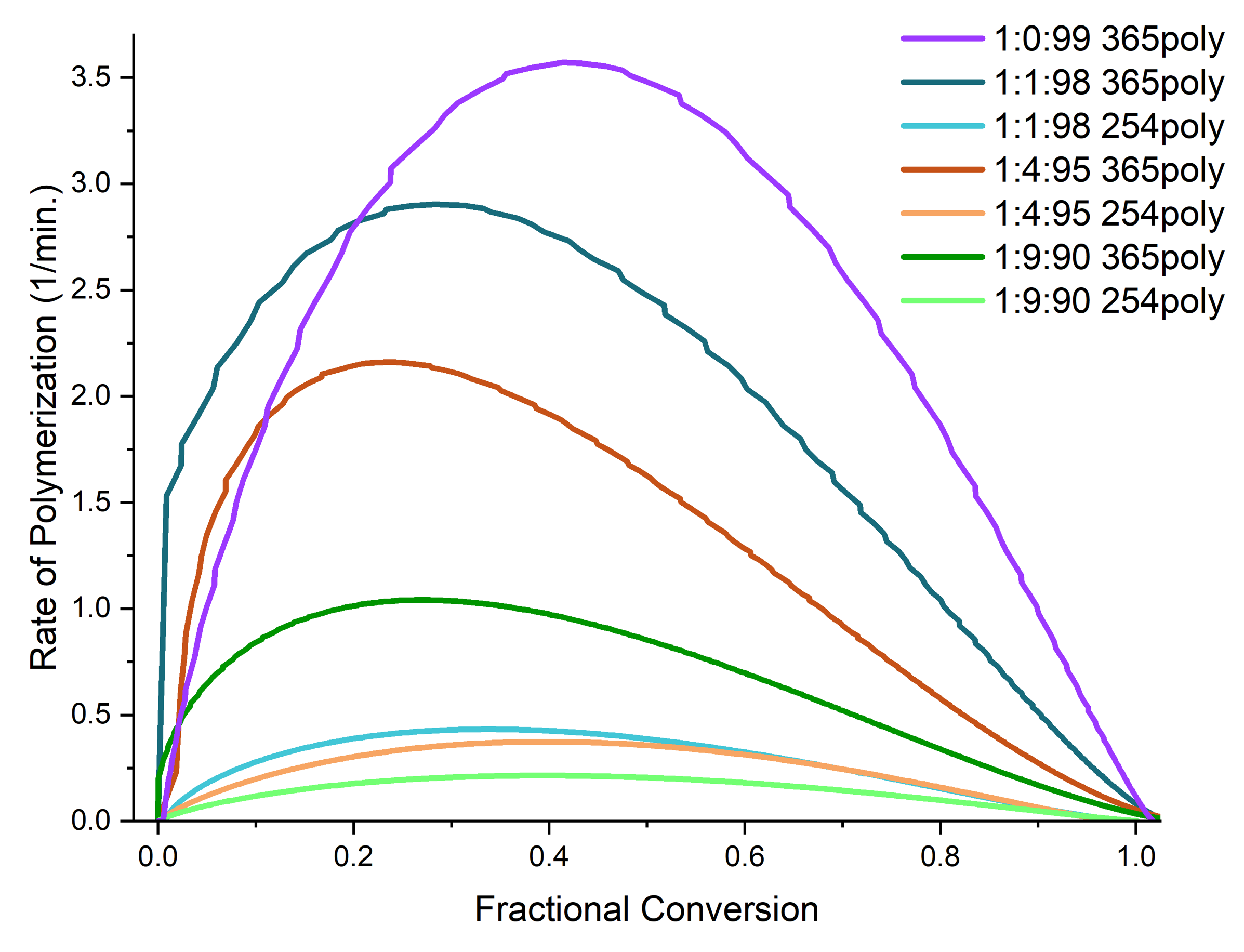}
    \caption{\textbf{Rate of polymerization}, calculated as the first derivative of fractional conversion versus time, is plotted versus fractional conversion for various PEGCoumAc formulation and polymerization conditions.}
    \label{fig:Rp vs conv}
\end{figure}

\begin{table}[H]
    \centering
        \caption{\textbf{The storage modulus at 26$\degree C$} values for all PEGCoumAc and some PEG formulations are included for both the 365 nm and 254 nm polymerization conditions.}
    \begin{tabular}{|c|c|c|}
    \cline{1-3}
    \multicolumn{3}{|c|}{Photo-polymerized at 365 nm}\\
    \cline{1-3}
    PEGDA (mol\%) & CoumAc (mol\%) & Storage Modulus at 26\degree C (MPa)\\
    \cline{1-3}
    0.5 & 0.5 & 0.044$\pm$0.0015\\
    \cline{1-3}
    0.5 & 4.5 & 0.0677$\pm$0.0084\\
    \cline{1-3}
    1 & 0 & 0.2252$\pm$0.194 \\
    \cline{1-3}
    1 & 1 & 0.0938$\pm$0.0039\\
    \cline{1-3}
    1 & 4 & 0.1332$\pm$0.0129\\
    \cline{1-3}
    1 & 9 & 0.1335$\pm$0.0039\\
    \cline{1-3}
    5 & 0 & 0.9491$\pm$0.0986\\
    \cline{1-3}
    5 & 5 & 1.0981$\pm$0.1203\\
    \cline{1-3}
    5 & 15 & 0.9597$\pm$0.0109\\
    \cline{1-3}
    10 & 0 & 1.803$\pm$0.0901\\
    \cline{1-3}
    \multicolumn{3}{|c|}{Photo-polymerized at 254 nm}\\
    \cline{1-3}
    PEGDA (mol\%) & CoumAc (mol\%) & Storage Modulus at 26\degree C (MPa)\\
    \cline{1-3}
    0.5 & 0.5 & 0.0868$\pm$0.0082\\
    \cline{1-3}
    0.5 & 4.5 & 0.1039$\pm$0.0043\\
    \cline{1-3}
    1 & 1 & 0.1625$\pm$0.0076\\
    \cline{1-3}
    1 & 4 & 0.1748$\pm$0.0046\\
    \cline{1-3}
    1 & 9 & 0.2805$\pm$0.1110\\
    \cline{1-3}
    5 & 5 & 0.9625$\pm$0.0461\\
    \cline{1-3}
    5 & 15 & 1.2753$\pm$0.0522\\
    \cline{1-3}
    \end{tabular}
    \label{tab:sm26C}
\end{table}

\begin{table}[H]
    \centering
        \caption\textbf{{Intensity measurements after light passes through a glass slide, a PEGCoumAc sample, and a glass slide and PEGCoumAc sample together.}}
    \begin{tabular}{|c|c|c|}
     \hline
     & \begin{tabular}{c}254 nm UV oven\\intensity ($\frac{W}{cm^2}$)\end{tabular}
     & \begin{tabular}{c}365 nm UV lamp\\intensity ($\frac{W}{cm^2}$)\end{tabular} \\
    \hline
        Raw measurement (no interference) & $8.5E10^{-3}$ & 0.1\\
        \hline
        Through glass & $1E10^{-4}$ & 0.09 \\
        \hline
        Through 1:9:90 PEGCoumAc sample (2 mm) & $1.9E10^{-4}$ & 0.02\\
        \hline
        Through glass and sample & $1.5E10^{-4}$ & 0.02\\
        \hline
    \end{tabular}
    \label{tab:intensitydrop}
\end{table}

\begin{figure}[H]
    \centering
    \includegraphics[width=0.8\linewidth]{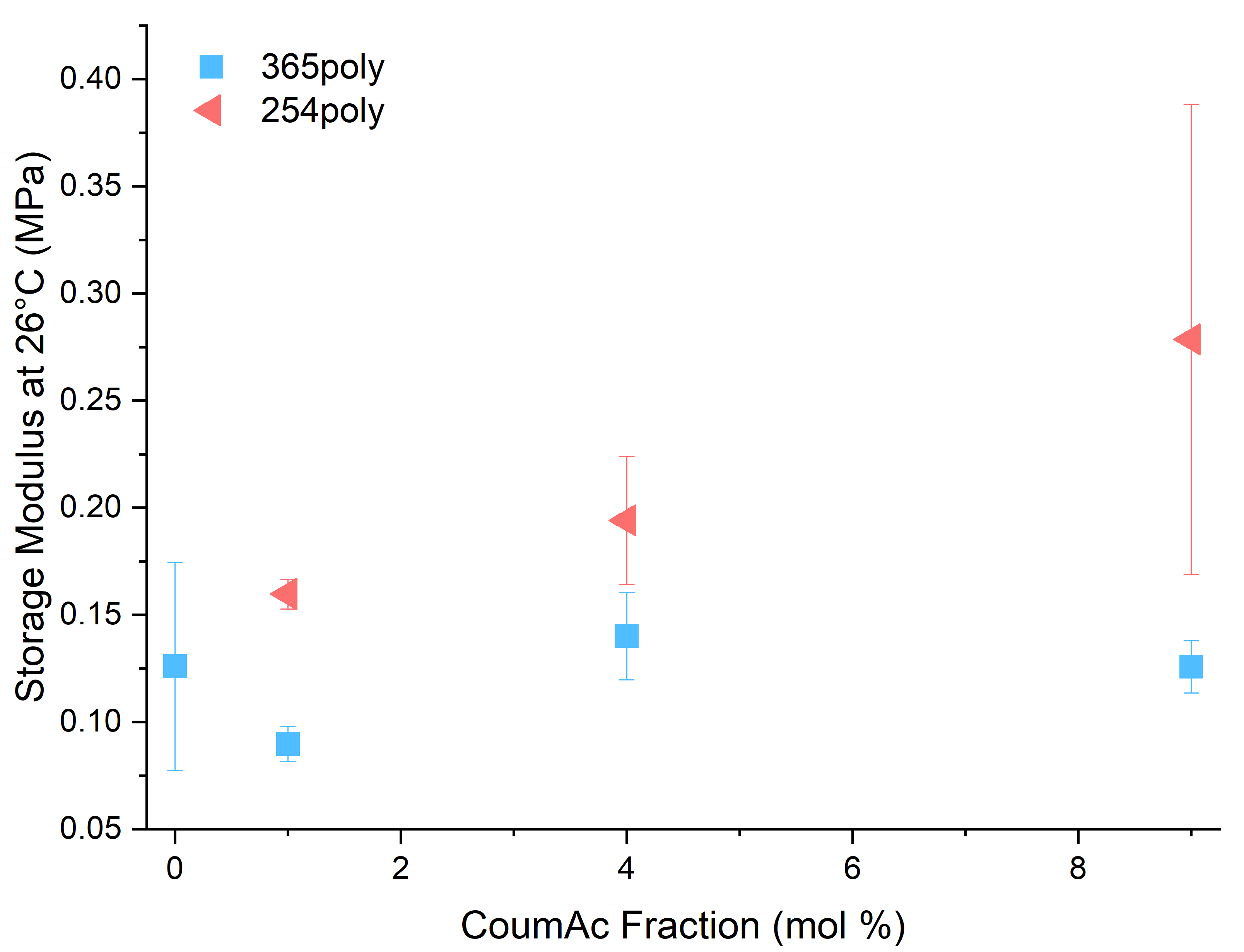}
    \caption{\textbf{Comparing the mechanical behavior of PEGCoumAc samples containing 1 mol\% PEGDA cured at either 365 or 254 nm are compared to the 1:99 base PEG system.} Storage modulus data measured \textit{via} DMA show that, at 26\degree C,  polymerizing at 254 nm, as opposed to 365 nm, and increasing CoumAc fraction both increase bulk stiffness.}
    \label{fig:storage modulus at room temp}
\end{figure}

\end{document}